\documentclass[camera,letterpaper,nomarginnotes,nonarrowgutter]{jpaper} 

\newif\ifcameraready
\camerareadytrue

\usepackage{cite}
\usepackage{amsmath,amssymb,amsfonts}
\usepackage[linesnumbered,ruled]{algorithm2e}
\usepackage{algorithmic}
\usepackage{datetime} 
\usepackage{graphicx}
\usepackage{textcomp}
\usepackage{xcolor}
\usepackage{fancyhdr}

\usepackage{todonotes}
\usepackage{xspace}
\usepackage[normalem]{ulem}
\usepackage{booktabs}
\usepackage{multirow}
\usepackage{setspace}
\usepackage{siunitx}
\DeclareSIUnit\bit{b}
\usepackage{soul}
\usepackage{pifont}

\usepackage{lipsum}
\usepackage{makecell}
\newcommand{\stripe}{\rowcolor{blue!5}}
\newcommand{\mr}[2]{\multicolumn{1}{c}{\multirow{#1}{*}{\makecell{#2}}}}
\usepackage{rotating}
\usepackage{tablefootnote}

\usepackage[bookmarks=true,breaklinks=true,letterpaper=true,colorlinks,linkcolor=black,citecolor=black,urlcolor=black]{hyperref}

\usepackage{titlesec}
\titlespacing\section{0pt}{*0.5}{*0.25}
\titlespacing\subsection{0pt}{*0.5}{*0.25}
\titlespacing\subsubsection{0pt}{*0.5}{*0.25}

\makeatletter
\g@addto@macro{\normalsize}{%
  \setlength{\abovedisplayskip}{2pt plus 1pt minus 1pt}
  \setlength{\belowdisplayskip}{2pt plus 1pt minus 1pt}
  \setlength{\abovedisplayshortskip}{0pt}
  \setlength{\belowdisplayshortskip}{0pt}
  \setlength{\intextsep}{2pt plus 1pt minus 1pt}
  \setlength{\textfloatsep}{4pt plus 1pt minus 1pt}
  \setlength{\skip\footins}{5pt plus 1pt minus 1pt}}
\makeatother

\expandafter\def\expandafter\UrlBreaks\expandafter{\UrlBreaks
  \do\a\do\b\do\c\do\d\do\e\do\f\do\g\do\h\do\i\do\j
  \do\k\do\l\do\m\do\n\do\o\do\p\do\q\do\r\do\s\do\t
  \do\u\do\v\do\w\do\x\do\y\do\z\do\A\do\B\do\C\do\D
  \do\E\do\F\do\G\do\H\do\I\do\J\do\K\do\L\do\M\do\N
  \do\O\do\P\do\Q\do\R\do\S\do\T\do\U\do\V\do\W\do\X
  \do\Y\do\Z\do\-}

\newcommand{\floor}[1]{\left\lfloor #1 \right\rfloor}

\newcommand{\boldone}{\ding{202}}
\newcommand{\boldtwo}{\ding{203}}
\newcommand{\boldthree}{\ding{204}}
\newcommand{\boldfour}{\ding{205}}

\newcommand{\One}{\emph{(i)}~}
\newcommand{\Two}{\emph{(ii)}~}
\newcommand{\Three}{\emph{(iii)}~}
\newcommand{\Four}{\emph{(iv)}~}
\newcommand{\Five}{\emph{(v)}~}

\newcommand{\trcd}{$t_{RCD}$}
\newcommand{\trp}{$t_{RP}$}
\newcommand{\tras}{$t_{RAS}$}

\newcommand{\X}[0]{QUAC\xspace}
\newcommand{\XT}[0]{QUAC-TRNG\xspace}
\newcommand{\nochips}[0]{136\xspace}
\newcommand{\nodimms}[0]{17\xspace}

\usepackage{tikz}
\newcommand*\hcircled[1]{\tikz[baseline=(char.base)]{\node[shape=circle, draw=black, fill=white, inner sep=.7pt] (char) {\textcolor{black}{#1}};}}

\ifcameraready

\definecolor{cerulean}{rgb}{0.0, 0.48, 0.65}

\newcommand{\agycomment}[1]{}

\definecolor{dollarbill}{rgb}{0.52, 0.73, 0.4}
\definecolor{ups-truck}{rgb}{0.53, 0.28, 0.21}
\newcommand{\taberk}[1]{\textcolor{black}{#1}}
\newcommand{\taberkcomment}[1]{}
\newcommand{\atbc}[1]{}

\definecolor{Moegi}{rgb}{0.357, 0.537, 0.188}

\newcommand{\hluocomment}[1]{}

\definecolor{darkamber}{rgb}{0.9, 0.49, 0.0}
\definecolor{debianred}{rgb}{0.84, 0.04, 0.33}

\newcommand{\mpc}[1]{}

\definecolor{gfored}{rgb}{0.580, 0.050, 0.211}
\newcommand{\om}[1]{\textcolor{black}{#1}}

\newcommand{\omc}[1]{}
\newcommand{\mc}[1]{}

\else

\definecolor{cerulean}{rgb}{0.0, 0.48, 0.65}

\newcommand{\agycomment}[1]{\textcolor{cerulean}{\textbf{\emph{[@gy: #1]}}}}

\definecolor{dollarbill}{rgb}{0.52, 0.73, 0.4}
\definecolor{ups-truck}{rgb}{0.53, 0.28, 0.21}
\newcommand{\taberk}[1]{\textcolor{ups-truck}{#1}}
\newcommand{\taberkcomment}[1]{\textcolor{ups-truck}{\textbf{[@atb: #1]}}}
\newcommand{\atbc}[1]{\todo[size=\scriptsize,color=ups-truck]{\textbf{Ata:} #1}}

\definecolor{Moegi}{rgb}{0.357, 0.537, 0.188}

\newcommand{\hluocomment}[1]{\textcolor{Moegi}{\textbf{[@hluo: #1]}}}

\definecolor{darkamber}{rgb}{0.9, 0.49, 0.0}
\definecolor{debianred}{rgb}{0.84, 0.04, 0.33}

\newcommand{\mpc}[1]{\textcolor{darkamber}{\textbf{[@minp: #1]}}}

\definecolor{gfored}{rgb}{0.580, 0.050, 0.211}
\newcommand{\om}[1]{\textcolor{blue}{#1}}

\newcommand{\omc}[1]{\todo[size=\scriptsize,color=gfored]{\textbf{Onur says:} #1}}
\newcommand{\mc}[1]{\todo[size=\scriptsize,color=darkamber]{\textbf{mpatel says:} #1}}
\fi

\newcommand{\affilETH}[0]{\textsuperscript{\S}}
\newcommand{\affilETU}[0]{\textsuperscript{$\dagger$}}
\newcommand{\affilUT}[0]{\textsuperscript{$\odot$}}

\begin{document}
\bstctlcite{IEEEexample:BSTcontrol}

\date{}
\title{\Large \XT: High-Throughput True Random Number Generation\\Using Quadruple Row Activation in Commodity DRAM {Chips}\vspace{-15pt}}
\author{
{Ataberk Olgun\affilETH\affilETU}\qquad%
{Minesh Patel\affilETH}\qquad
{A. Giray Ya\u{g}l{\i}k\c{c}{\i}\affilETH}\qquad
{Haocong Luo\affilETH}\qquad \\%
{Jeremie S. Kim\affilETH}\qquad%
{F. Nisa Bostanc{\i}\affilETH\affilETU}\qquad
{Nandita Vijaykumar\affilETH\affilUT}\qquad%
{O\u{g}uz Ergin\affilETU}\qquad%
{Onur Mutlu\affilETH}\qquad\vspace{-3mm}\\\\
{\vspace{-3mm}\affilETH \emph{ETH Z{\"u}rich}} \qquad \affilETU \emph{TOBB University of Economics and Technology} \qquad \affilUT \emph{University of Toronto} %
}

\maketitle

\def\parsepdfdatetime#1:#2#3#4#5#6#7#8#9{%
  \def\theyear{#2#3#4#5}%
  \def\themonth{#6#7}%
  \def\theday{#8#9}%
  \parsepdftime
}

\def\parsepdftime#1#2#3#4#5#6#7\endparsepdfdatetime{%
  \def\thehour{#1#2}%
  \def\theminute{#3#4}%
  \def\thesecond{#5#6}%
  \ifstrequal{#7}{Z}
  {%
    \def\thetimezonehour{+00}%
    \def\thetimezoneminute{00}%
  }%
  {%
    \parsepdftimezone#7%
  }%
}

\def\parsepdftimezone#1'#2'{%
  \def\thetimezonehour{#1}%
  \def\thetimezoneminute{#2}%
}

\newcommand*{\thetimezone}{\thetimezonehour:\thetimezoneminute}
\expandafter\parsepdfdatetime\pdfcreationdate\endparsepdfdatetime

\settimeformat{ampmtime}
\newcommand{\versionnum}[0]{8.0 - Final}
\newcommand{\version}[1]{\emph{Version #1 (Built:~\today~@ \currenttime~UTC\thetimezone)}}

\fancyhead{}
\ifcameraready
\thispagestyle{empty}
\pagestyle{empty}
\else
\fancyhead[C]{\textcolor{blue}{\version{\versionnum}}}
\fancypagestyle{firststyle}
{
  \fancyhead[C]{\textcolor{blue}{\version{\versionnum}}}
  \fancyfoot[C]{\thepage}
}
\thispagestyle{firststyle}
\pagestyle{firststyle}
\fi


\sloppypar

\begin{abstract}
{True random number generators (TRNG) sample random physical processes to create large amounts of random numbers for various {use cases,} including security-critical cryptographic primitives, scientific simulations, machine learning applications, and even recreational entertainment. Unfortunately, not every computing system is equipped with dedicated TRNG hardware, limiting the application space and security guarantees for such systems. To open the application space and enable security guarantees {for} the overwhelming majority of computing systems that do not necessarily have dedicated TRNG hardware {(e.g., processing-in-memory systems)}, we develop \XT, a {new} high-throughput TRNG that can be fully implemented in commodity DRAM chips, which are key {components in} most modern systems.} 

{QUAC-TRNG exploits the {new} observation that a carefully-engineered sequence of DRAM commands activates four consecutive DRAM rows in rapid succession. This QUadruple ACtivation (QUAC) {causes} the bitline sense amplifiers {to} non-deterministically converg{e} to random values when {we} activat{e four} rows that store conflicting data because the net deviation in bitline voltage fails to meet reliable sensing margins.} 

{We experimentally demonstrate that \X reliably generates random values across \nochips commodity DDR4 DRAM chips from one major DRAM manufacturer.}
{We describe how to develop an effective TRNG (\XT) based on \X. We evaluate the quality of our TRNG using the commonly-used NIST statistical test suite for randomness and find that \XT successfully passes each test.} {{Our experimental} evaluat{ions show that} \XT reliably generates true random numbers with {a throughput of {3.44} Gb/s} {(per DRAM channel)}}{, outperforming the state-of-the-art DRAM-based TRNG by {{15.08}$\times$} and {1.41}$\times$ for bas{ic} and throughput-optimized versions, respectively.}
{{{W}e} show that \XT utilizes DRAM bandwidth {better than} the state{-}of{-}the{-}art,} {{achieving} up to {2.03}$\times$ {the throughput of a} throughput-optimized baseline when scaling bus frequencies to 12 GT/s.}

\end{abstract}

\setstretch{0.92}

\section{Introduction}
\label{sec:introduction}
{True random numbers are used in a wide range of applications\taberk{,} including cryptography, scientific simulations, machine learning, and recreational {entertainment}~\cite{darrell1998genetic, vincent2010stacked, schmidt1992feedforward, zhang2016survey, mich2010machine, bagini1999design, bakiri2018survey,
rock2005pseudorandom, ma2016quantum, stipvcevic2014true,
barangi2016straintronics, tao2017tvl, gutterman2006analysis, von2007dual,
kim2017nano, drutarovsky2007robust, kwok2006fpga, cherkaoui2013very,
zhang2017high, quintessence2015white}. These applications often require a high-throughput true random number generator (TRNG)} that is resilient to variations in operating conditions {(e.g., temperature and voltage fluctuations) and {is} secure against} malicious attacks~\cite{yang2016all}. 

{Unfortunately, {not all} computing systems are {provisioned with} dedicated TRNG hardware, limiting their ability to run such applications {effectively}. 
In order to address this issue, many works have attempted to provide true random number generators purely using commodity hardware components that can be found in most systems today (e.g., DRAM~\cite{kim2019drange,sutar2016d,talukder2019exploiting,keller2014dynamic,pyo2009dram} and SRAM\cite{ 
	  holcomb2007initial, holcomb2009power, 
	van2012efficient}).} 

{Using DRAM as the entropy source for generating true random numbers (i.e., DRAM-based TRNG) is a promising approach to providing a TRNG to} a variety of computing systems ranging from high-performance servers{,} low-power edge devices{, and systems that employ processing-in-memory}~\cite{upmem2018} {due to the widespread adoption of DRAM as main memory across these systems}. 
{However, prior proposals for DRAM-based TRNGs} {\One have high latencies in generating random numbers because they rely on fundamentally slow processes (e.g., retention failures~\cite{tehranipoor2016robust, keller2014dynamic, sutar2018d, hashemian2015robust}, DRAM start-up values~\cite{eckert2017drng}) or \Two generate random numbers at low throughput because they either use small portion{s} of {selected} DRAM rows as {an} entropy source (e.g., \trcd{} failure-based~\cite{kim2019drange}) or use whole DRAM rows as {an} entropy source but fail to induce metastability {in} many sense amplifiers (e.g., \trp{} failure-based~\cite{talukder2019exploiting}).}

\textbf{Our goal} in this work is to develop a TRNG that uses commodity DRAM devices to generate random numbers {with both high throughput and low latency. To achieve this, we leverage the novel observation that a carefully{-}engineered sequence of DRAM commands (described in Section~\ref{sec:quac}) activates four DRAM rows in quick succession in commodity DRAM chips from one major DRAM manufacturer {(SK {H}ynix)}, a process we refer to as \emph{QUadruple ACtivation} (\X).}

Our key idea is to leverage \X as a substrate for low-latency and high-throughput DRAM-based TRNGs. {When activating rows that are initialized with conflicting data (e.g., data `0' in two rows and data `1' in the other two), bitline sense amplifiers non-deterministically converge to random values based on their individual {circuit characteristics resulting from} {manufacturing} process variation.} {Using \X operations to induce metastability in many DRAM sense amplifiers in parallel enables high-throughput and low-latency random number generation.}

{To this end, we} develop \XT, a DRAM-based TRNG that repeatedly performs \X operations {in} DRAM and processes the results of these operations using {a cryptographic hash function~\cite{fips2012180}} to generate random numbers {with} high throughput. One \XT iteration consists of five key steps: \XT \One identifies four consecutive DRAM rows, \Two initializes the rows with {conflicting} data patterns {(e.g., data `0' in two rows and data `1' in the other two)}, \Three performs a \X operation on the rows by issuing a sequence of DRAM commands, \Four reads the result of the operation from {the sense amplifiers}, {and} \Five performs the SHA-256 cryptographic hash {function~\cite{fips2012180}} {to post-process} the result {and} output random numbers.
{Our experimental evaluation using {\nochips{} real DDR4 DRAM chips from \nodimms real DDR4 modules} (Section~\ref{sec:evaluation}) shows that \XT generates an average of {{7664}} bits of random data per iteration {and each iteration takes 1940 ns}.}

Compared to previously-proposed DRAM-based {T}RNGs~\cite{talukder2019exploiting,kim2019drange,pyo2009dram,sutar2016d,keller2014dynamic,eckert2017drng,humood2021DTRNG}, \XT enables {\One} lower latency because it only requires {simultaneous} activation of consecutive rows, which can be performed quickly using DRAM commands, and {\Two} higher throughput because it uses \X operations to induce metastability in many {sense amplifiers} in parallel.

{We evaluate \XT's quality by showing that random bitstreams generated using real DRAM chips pass the NIST statistical test suite~\cite{rukhin2001statistical} (Section~\ref{sec:quac-trng-output-eval}).}
{We then quantitatively evaluate \XT's performance against two state-of-the-art DRAM-based TRNG proposals~\cite{kim2019drange,talukder2019exploiting} (Section~\ref{sec:comparison-high-throughput}).
For each prior proposal, we consider two configurations: \One an unmodified \emph{base} version as proposed in the original paper and \Two an \emph{enhanced} version that we believe represents a more fair comparison against our work. The enhanced versions incorporate optimizations to improve throughput and employ} the SHA-256 hash function for post{-}processing. Our results show that \XT's throughput {is 15.08$\times$ and 1.41$\times$ that of} the {best prior DRAM-based TRNG for the bas{ic} and enhanced configurations, respectively.}
{{W}e show that \XT scales quasi-linearly with available DRAM bandwidth, outperforming} the \emph{enhanced} configuration of the {best prior} DRAM TRNG by up to \emph{{2.03}x} at future DRAM transfer rates. 
{{W}e {also} study {and demonstrate} how \XT can be integrated into a real system (Section~\ref{sec:integration}) with minor performance, memory capacity, and CPU die area costs.}

We make the following key contributions:
\begin{itemize}
    \item We make the novel observation that a carefully engineered sequence of DRAM commands can activate four DRAM rows in quick succession. We refer to this operation as QUadruple ACtivation (QUAC). 
    {We show that QUAC operations can induce metastability in DRAM bitline sense amplifiers, which we exploit to generate true random numbers.}
    \item {We introduce \XT{}, a new high-throughput TRNG based on \X operations that is suitable for commodity DRAM chips. \XT{} {combines the benefits of two components to generate high-quality true random numbers with high throughput: (i)} massive parallelism in {true random number generation available in DRAM sense amplifiers}} {and (ii) {{randomness quality {improvements} provided by} the SHA-256 hash function}} to generate random numbers at significantly higher throughput than previously-proposed DRAM-based TRNGs. 
    
    \item 
    {We experimentally demonstrate that \XT{} {is a high-quality TRNG} {by showing that the random bitstreams \XT generates pass \emph{all} {the standard} NIST {statistical test suite}}
    {randomness tests~\cite{rukhin2001statistical}}.}
    
    \item
    {We show that \XT{} improves throughput over state-of-the-art DRAM-based TRNG proposals{~\cite{talukder2019exploiting,kim2019drange}}, achieving 15.08$\times$ and 1.41$\times$ {the} throughput {of} basic and throughput-optimized baselines, respectively.}
    
    \item
    {We present a detailed {experimental} characterization study o{f} the randomness {provided by} \X operations using {\nochips{} real DDR4 chips (from \nodimms{} DDR4 modules)}. {We} {show} that {\One} \XT is suitable for implementation in commodity DRAM chips, and \Two the randomness {provided} by \X operations remain{s} stable over time.}
  
    \end{itemize}

\section{Background}
\label{sec:background}
\subsection{DRAM Structure and Organization}
\label{sec:background-dram}

DRAM-based main memory is organized hierarchically. A processor is connected to one or many \textit{memory channels}. Each channel has its own command, address, and data buses. Multiple \textit{memory modules} can be plugged into a single channel. Each module {contains several \textit{DRAM chips}, which are grouped into} \textit{ranks}. Each rank contains multiple \textit{banks} that are {striped across the chips that form the rank but} operate independently. {Particular standards cluster m}ultiple \textit{banks} in \textit{bank groups}~\cite{jedecDDR4,gddr5}.
{D}ata transfers between DRAM memory modules and processors {occur at a} \textit{cache block} granularity. 

\begin{figure}[!ht]
\centering
\includegraphics[width=0.49\textwidth]{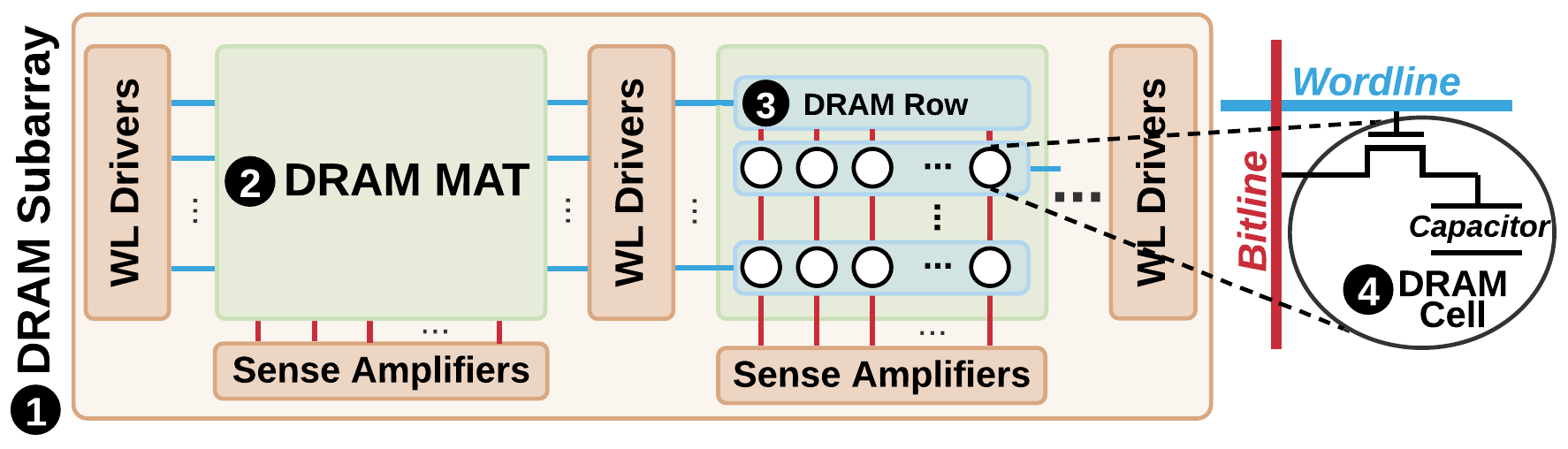}
\caption{DRAM subarray{, MAT{, row} and cell organization}}
\label{fig:bank_organization}
\end{figure}

\noindent
\textbf{DRAM Bank Organization.}
\textit{A DRAM bank} is divided into multiple \textit{subarrays}~\cite{kim2012case,chang2014improving,seshadri2020indram}{. Each subarray comprises multiple wordline drivers and sense amplifiers (SAs)}{,} as {shown} in Figure~\ref{fig:bank_organization}{-\boldone{}{.}} {Subarrays are further divided into \emph{DRAM MATs}. Figure~\ref{fig:bank_organization}-\boldtwo{} shows a DRAM MAT.} DRAM MATs are separated {from each other} by \emph{wordline {(WL)} drivers} that are activated to drive {a} DRAM wordline {within the DRAM MAT}. {In a DRAM MAT, \emph{DRAM cells} are organized into a two{-}dimensional structure over \emph{bitlines} and \emph{wordlines}. The set of cells over the same wordline forms a \emph{DRAM row}, as depicted in Figure~\ref{fig:bank_organization}-\boldthree.}

\noindent
\textbf{Accessing DRAM.}
A DRAM cell {(Figure~\ref{fig:bank_organization}-\boldfour{})} stores data as {a voltage level between the supply voltage ($V_{DD}$) and ground} in {its} capacitor. Each cell is connected to a bitline via an \textit{access transistor}. When all rows are closed, bitlines are precharged to {the half of supply voltage ($V_{DD}/2$)}. {Accessing a cell requires activating the corresponding row by issuing an {(\emph{ACT})} command. The activation process starts with enabling a wordline, {which} enabl{es} all access transistors in the row. As the access transistors are turned on, each cell shares its charge with the corresponding bitline{, causing} deviat{ion on} the bitline voltage either towards $V_{DD}$ or ground.}
Each SA amplifies a bitline's voltage to either $V_{DD}$ or $0$ as the deviation in bitline voltage exceeds {a threshold voltage ($V_{th}$)}. Read and write operations can be issued to {SAs {only} after the row activation is completed}. {A} precharge {(\emph{PRE})} command {is} used to close a row and set the bitline voltage to $V_{DD}/2$.

\begin{figure}[!ht]
\centering
\includegraphics[width=0.49\textwidth]{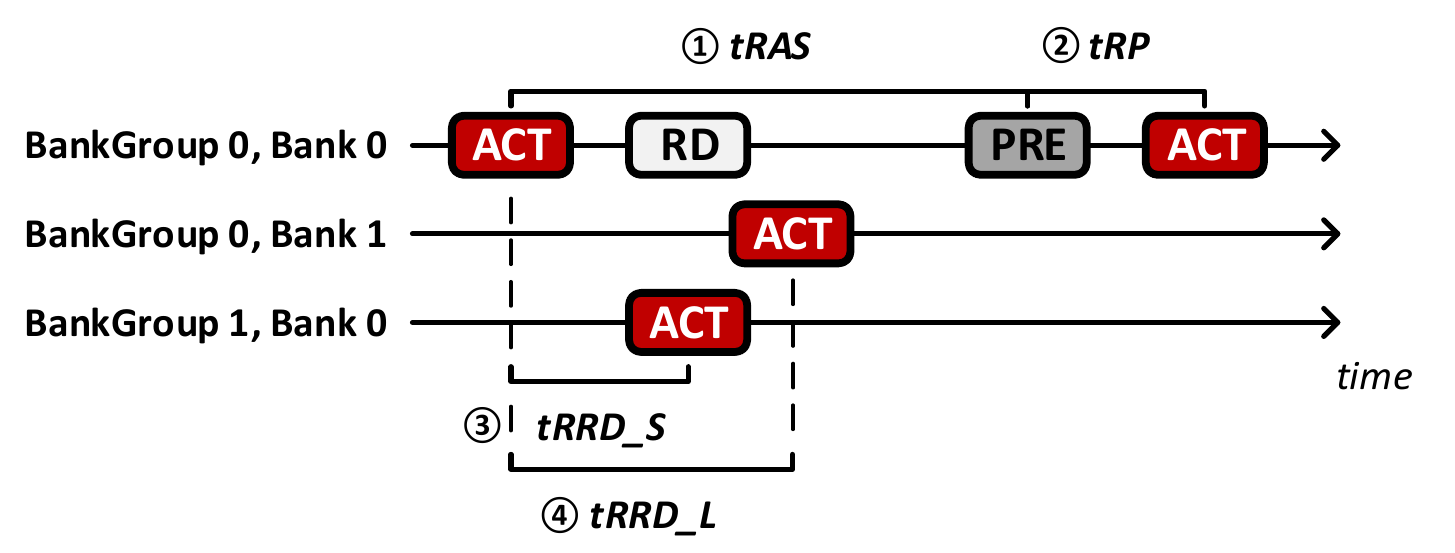}
\vspace{-4.5mm}
\caption{Timeline of {key} DDR4 commands.}
\label{fig:ddr-timings}
\end{figure}

\noindent
\textbf{DRAM Timing Parameters.}
{A memory controller} must obey the DRAM timing parameters defined in {standards set by JEDEC (e.g., DDR4}~\cite{jedecDDR4}) {while scheduling DRAM commands}. Figure~\ref{fig:ddr-timings} presents a timeline of DRAM commands on the command bus.
{Consecutive {\emph{ACT}} and {\emph{PRE}} commands on the command bus must be interleaved by {at least} \tras{} {(i.e., ACT $\rightarrow{}$ {PRE} timing parameter)} {(\boldone{})}. {This is because} a row needs to be active for at least as long as \tras{} {to allow its cells to fully restore their charge{.}}}
{The time window between a {\emph{PRE}} and {an} {\emph{ACT}} command on the command bus must be {at least} $t_{RP}$ {(\boldtwo{})}. This is required to {settle} the bitline voltage to $V_{DD}/2$ and to disable the activated wordline.} 
Back-to-back {\emph{ACT}} commands to {different bank groups and to} different DRAM banks {(in the same bank group)} must be interleaved {with latencies of} {\emph{$t_{RRD\_S}$}} {(\boldthree{})} and {\emph{$t_{RRD\_L}$}} {(\boldfour{}),} respectively. {\emph{$t_{RRD\_S}$}} and {\emph{$t_{RRD\_L}$}} are usually small (3.00, 4.90 ns in DDR4-2666{~\cite{jedecDDR4}}). This enables overlapping the activation latency of DRAM rows in different banks or bank groups.

DRAM manufacturers set large guardbands around DRAM timing parameters to guarantee correct operation~\cite{aldram, chang2016understanding, chang_understanding2017, lee2017design, kim2018solar}. A large body of work characterizes DRAM behavior under non-standard DRAM timing parameters to demonstrate that {violating DRAM timing parameters allows improving} DRAM access latency~\cite{aldram, chandrasekar2014exploiting,  chang2016understanding, lee2016phdthesis, lee2016reducing, lee2017design, chang_understanding2017, chang2017thesis, kim2018solar, kim2020phdthesis}, generating random numbers~\cite{kim2019drange, talukder2019exploiting, kim2020phdthesis}, implementing physical unclonable functions (PUFs)~\cite{talukder2019exploiting,kim2018latpuf, kim2020phdthesis}, {and copying data and performing} bitwise AND/OR in DRAM~\cite{fei2019computedram} on commodity DRAM devices.

\subsection{True Random Number Generators}
\label{sec:background-rng}

True random number generators (TRNGs){~\cite{kocc2009cryptographic}} harness entropy from random physical phenomena to generate random numbers. These entropy sources are often biased~\cite{kocc2009cryptographic, stipvcevic2014true}, {so practical TRNG designs often} use post-processing methods to remove bias in their entropy sources, i.e., to strengthen the quality of the random numbers they produce {(e.g., hashing~\cite{rivest1992md5} and other whitening algorithms~\cite{jun1999intel, von2007dual}). {Post-processing can constrain TRNG throughput and latency, potentially requiring additional resources (e.g., output buffering) to offset its impact.}} 

\section{Motivation and Goal}
\label{sec:motivation}
High-quality random numbers are {crucial} to many technologies and applications~\cite{gutterman2006analysis, von2007dual, kim2017nano, drutarovsky2007robust, kwok2006fpga, cherkaoui2013very, zhang2017high, quintessence2015white,clarke2011robust,lu2015fpga,hull1962random, ma2016quantum, botha2005gammaray, davis1956some,zhang2016survey,mich2010machine,schmidt1992feedforward,darrell1998genetic,stipvcevic2014true}.
{In particular}, random numbers are used widely in cryptographic communication protocols (e.g.{,} key generation to initialize communication, signature and fingerprint generation to authenticate remote parties) to form secure channels between computing systems and {networked devices}. These protocols \emph{require} an unpredictable, high-quality stream of \emph{true random numbers} to remain secure against cryptographic attacks~\cite{vincent2010stacked, cherkaoui2013very} that aim to breach highly valuable, confidential user data. Some emerging key distribution protocols (e.g., quantum key distribution) provide even {stronger} security guarantees that make them resilient against a more diverse set of attacks~\cite{clarke2011robust,lu2015fpga}. These protocols require TRNG throughputs {on the order of several} Gb/s~\cite{wang2016theory}. 
Other than cryptography, high-throughput TRNGs are useful for other applications such as scientific simulations~\cite{hull1962random, ma2016quantum, botha2005gammaray, davis1956some}, machine learning~\cite{zhang2016survey,mich2010machine,schmidt1992feedforward,darrell1998genetic}, and gaming applications~\cite{stipvcevic2014true}{.}

\noindent
\textbf{High-throughput TRNGs.} 
Many prior {works develop and demonstrate high-throughput TRNGs that}
use specialized hardware (e.g., optics~\cite{zhang2017gbit, gehring2020ultra, ugajin2017real,stefanov2000optical}, ring oscillators~\cite{wang2016gbps,amaki2015oscillator,yang2016all,bucci2003high}, chaotic circuits~\cite{pareschi2006fast,restituto1993nonlinear}) to generate random numbers. {Unfortunately, these proposals typically either \One need to be integrated at design time, rendering them unsuitable for existing systems or \Two are costly, limiting their potential for widespread adoption.
To overcome these limitations and enable the aforementioned applications across computing systems ranging from high-performance servers to low-power edge devices, it is important to enable high-quality random number generation using existing commodity hardware.}

\noindent
\textbf{DRAM-based TRNGs.} {DRAM is a promising substrate for true random number generation because DRAM chips are ubiquitous throughout contemporary computing platforms. DRAM-based TRNGs can be integrated into commodity systems {at low cost} with {minimal} effort~\cite{kim2019drange}, thereby enabling high-throughput random number generation across a broad spectrum of both \One existing and \Two future computing systems.}

\noindent
\textbf{Synergy With PIM.} Processing-in-memory (PIM) systems improve system performance and/or energy consumption by performing computations directly within a memory chip, thereby avoiding unnecessary data movement~\cite{seshadri2017ambit, ghose2018enabling, seshadri2017simple, oliveira2021new, juan2021benchmarking, boroumand2021polynesia, boroumand2019conda, mutlu2020modern, ghose2019processing}. Prior works propose a broad range of PIM systems~\cite{wang2020figaro,giannoula2021syncron,akin2015data, aga2017compute, ahn2015scalable, ahn2015pim, lee2015simultaneous, seshadri2015fast, seshadri2013rowclone, seshadri2015gather, seshadri2017ambit, liu2017concurrent, seshadri2017simple, pattnaik2016scheduling, babarinsa2015jafar, farmahini2015nda, gao2015practical, gao2016hrl, hassan2015near, hsieh2016transparent, morad2015gp, sura2015data, zhang2014top, hsieh2016accelerating, boroumand2017lazypim, chang2016low, kim2018grim, ghose2018enabling, boroumand2021mitigating, boroumand2018google, seshadri2016simple, mutlu2019processing, CROW, singh2020near, fernandez2020natsa, kwon202125, devaux2019true, li2016pinatubo, chi2016prime, orosa2019dataplant, orosa2021codic} in the context of various workloads and memory devices. 
{Enabling new PIM workloads (e.g., security applications) that rely on high-quality random numbers requires allowing the PIM system to perform TRNG operations directly within the memory to both (1)} avoid inefficient off-chip communication to other possible TRNG sources, {and (2)} to enhance the overall security and privacy of PIM systems.

\noindent
\textbf{Shortcomings of Prior Work.} Prior proposals for DRAM-based TRNGs {either} \One have high latencies in generating random numbers because they rely on fundamentally slow processes (e.g., retention failures~\cite{tehranipoor2016robust, keller2014dynamic, sutar2018d, hashemian2015robust}, DRAM start-up values~\cite{eckert2017drng}) {or} \Two generate random numbers at low throughput because they either use small portion{s} of {selected} DRAM rows as entropy source (e.g., \trcd{} failure-based~\cite{kim2019drange}) or use whole DRAM rows as entropy source but fail to induce metastability on many sense amplifiers (e.g., \trp{} failure-based~\cite{talukder2019exploiting}).

{TRNGs based on DRAM start-up values}{~\cite{eckert2017drng}} require a power cycle to generate random bits. This mechanism is impractical for a high-throughput TRNG {because it both} \One{} {incurs} very high random number generation latency and \Two{} {precludes generating} random bits in a streaming manner. {TRNGs based on DRAM retention failures~\cite{keller2014dynamic,sutar2016d}} need to accumulate DRAM retention failures over long periods of time to harness enough entropy to generate random numbers. DRAM cells flip very infrequently due to retention failures as \om{many} DRAM cells retain data \om{for hours}~\cite{khan2014efficacy,venkatesan2006retention, patel2017reaper, raidr, qureshi2015avatar}. The throughput of activation latency-based TRNGs{~\cite{talukder2019exploiting,kim2019drange}} is constrained by the amount of entropy they can harness from small portion{s} of {selected} DRAM rows, a DRAM cache block. {For example,} D-RaNGe~\cite{kim2019drange} can only use up to \emph{4} out of the \emph{64K} bits available for random number generation. Precharge latency-based TRNGs induce bit-flips on many DRAM cells in parallel on DRAM row granularity. However, the proportion of randomly-failing cells among all cells in a DRAM row following precharge latency failures is very low.\footnote{{{Section~\ref{sec:comparison-high-throughput}} provides a rigorous analysis of prior DRAM-based TRNGs}}

{We posit from our analysis o{f} prior work that a high-throughput DRAM-based TRNG needs to \One {exploit} DRAM failure mechanisms that are inherently fast and random (e.g., timing failures), \Two harness entropy from large portions of selected DRAM rows, and \Three induce random behavior on a large proportion of sense amplifiers.} 

\textbf{Our goal} is to develop a {new TRNG mechanism} that uses commodity DRAM devices to {robustly} generate {high-quality} random numbers with high{er} throughput and low latency. 

\section{Quadruple Activation}
\label{sec:quac}
{{We observe a new phenomenon, which we call} \textbf{qu}adruple \textbf{ac}tivation (QUAC){,}} {in commodity DRAM modules. We find that by issuing a sequence of ~{three standard DDR4 commands} \emph{(ACT $\rightarrow{}$ PRE $\rightarrow{}$ ACT)} with {reduced timings} (e.g., 2.5 ns), four consecutive DRAM rows in the same subarray are activated simultaneously. We identify the following two characteristics of \X{}.} 
{First, \X{} can simultaneously activate a set of four DRAM rows whose row addresses differ \emph{only} in their two least significant bits (e.g., rows \{0,1,2,3\}). We refer to each such set of four DRAM rows as a \emph{DRAM segment}.}
{Second, {we observe }{QUAC only {when} the two \emph{ACT} commands target row addresses whose two least significant bits are inverted.}
In other words, the two \emph{ACT} commands should target rows 0 and 3 (00 and 11 {in base 2}), or rows 1 {and} 2 (01 and 10 {in base 2}) within a {DRAM} segment.}

{To explain the potential mechanism behind \X, we examine the array architecture in {state-of-the-art} high-density DRAM {chips}. We hypothesize that the hierarchical design of wordlines 
{allows \X to simultaneously activate four rows in a segment,}
and we present a hypothetical row decoder circuit that explains why the row addresses of the two \emph{ACT} commands must have their two {least} significant bits set to inverted values.
}

\subsection{Hierarchical Wordlines}
\begin{figure}[!ht]
\centering
\vspace{2mm}
\includegraphics[width=0.49\textwidth]{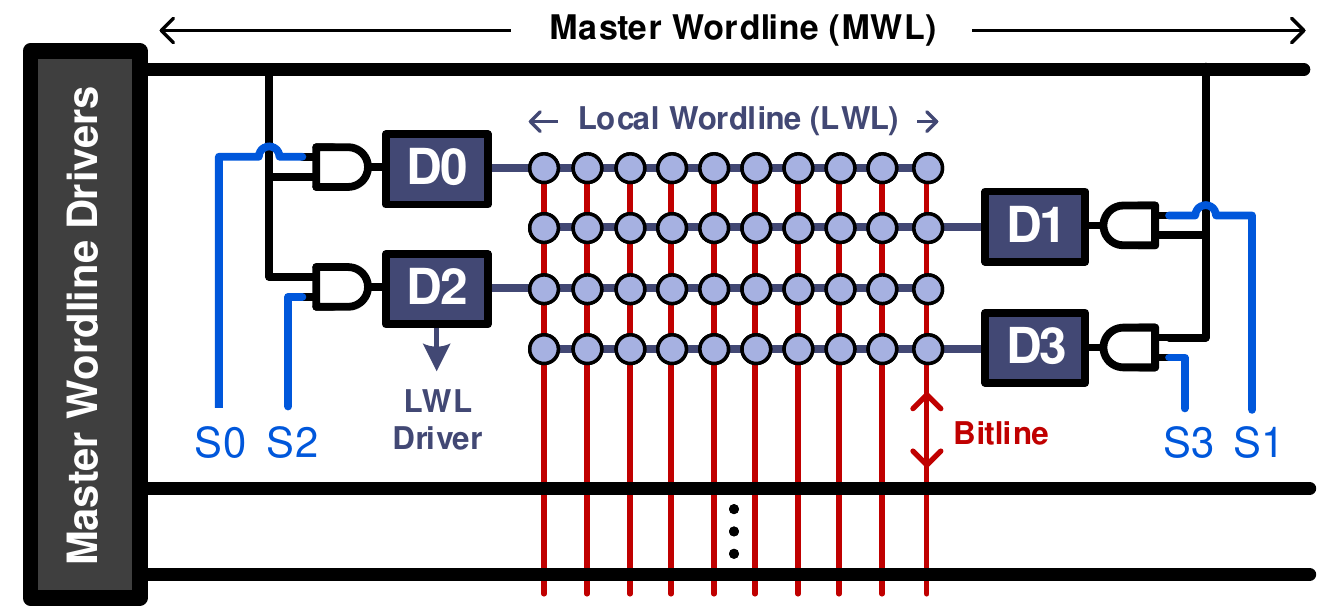}
\caption{DRAM MAT with hierarchical wordlines}
\vspace{2mm}
\label{fig:wordlines}
\end{figure}

High density and performance requirements have pushed DRAM designers to architect {high-density, low-latency} DRAM array architectures~\cite{mutlu2013memory}. {A commonly{-}used design pattern in architecting such DRAM arrays is to hierarchically organize DRAM wordlines to reduce latency and improve density~\cite{chatterjee2017architecting,udipi2010rethinking,rambusmodel,lee2017design}. Figure~\ref{fig:wordlines} shows a DRAM MAT with the hierarchical wordline design. }

{{In the hierarchical wordline design}, a DRAM row address is partitioned into two {pieces}. The higher-order bits {of the row address}
are used to select and activate a master wordline (MWL). {The MWL} is connected to four local wordline (LWL) drivers {(D0, D1, D2, D3 in Figure~\ref{fig:wordlines})} {that are used to activate} four consecutive DRAM rows in a MAT. The least significant two bits {of the row address} are used to assert one of the four LWL select lines (S0 to S3) to enable an LWL driver and finally activate a {DRAM row}.}

An activated MWL potentially drives four consecutive LWLs that form a segment. We hypothesize that the QUAC command sequence (ACT-PRE-ACT) asserts S0 to S3 {approximately} {at the same time, resulting in simultaneous activation of four consecutive DRAM rows.}

\subsection{Hypothetical Row Decoder}

{We present a hypothetical row decoder circuit design that supports \X{} operations. The decoder design simultaneously activates four DRAM rows when the DRAM chip receives a series of ACT-PRE-ACT commands with violated timing parameters.}
Figure 4 illustrates our row decoder circuit, which operates on the least significant two bits of row addresses. 

\begin{figure}[!ht]
\centering
\includegraphics[width=0.5\textwidth]{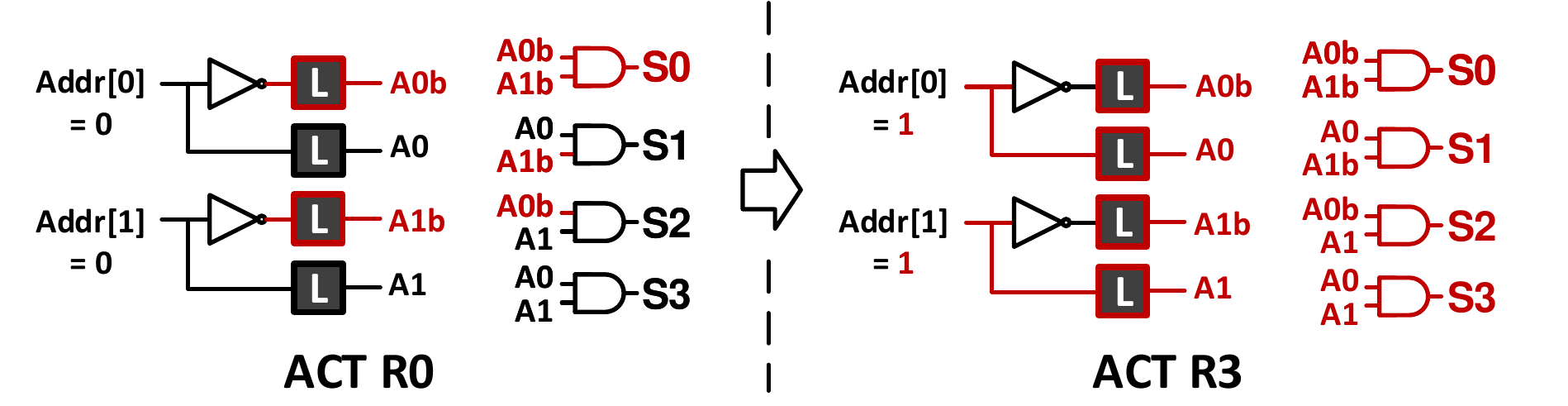}
\caption{Hypothetical row decoder circuit that enables \X. {The red and black colors represent asserted and de-asserted signals, respectively.}}
\label{fig:decoder}
\end{figure}

The first \emph{ACT} command {(Figure~\ref{fig:decoder}, left)} targeting {Row 0 (R0, Addr[1:0] = ``00'')} sets the latches {(L)} that drive the signals \emph{A0b} and \emph{A1b}. These signals are combined through a logical-AND operation to form S0, which enables the LWL driver that activates R0. The {following} \emph{PRE} command {\emph{cannot}} deactivate R0 nor reset the latches that drive \emph{A0b} and \emph{A1b}, as the \tras{} parameter is violated. The second \emph{ACT} command {(Figure~\ref{fig:decoder}, right)} targeting {Row 3} {({R3, }Addr[1:0] = ``11'')} sets {the} latches that drive the signals {\emph{A0}} and {\emph{A1}}. After the second \emph{ACT} command, all four control signals {(i.e., A0, A0b, A1, and A1b)} are enabled {since the previous \emph{PRE} command fails to reset the latches}. Together, these signals assert S1, S2, and S3, enabling the LWLs that activate R1, R2, and R3, respectively{. Since R0 is still activated, this results} in simultaneous activation of all four rows in a DRAM segment.

{We confirm that \X activates four DRAM rows through an experiment with real DRAM chips.} We first initialize a DRAM segment with a {predefined} data pattern. We then perform a \X operation on the DRAM segment to simultaneously activate four rows. Next, we write {a new data pattern} to the sense amplifiers while all four rows are active. Finally, we precharge the bank and {individually} read {each row} while obeying manufacturer-recommended DRAM timing parameters. We observe that all four rows are updated {with {the} new data {pattern we write}}.
{We observe valid \X operations in \nochips{} DDR4 chips from one {major DRAM} manufacturer.} 

{\subsection{Future \X{} Interfaces}}
{Even though current DDRX interfaces do not support QUAC, future DRAM chips can be built (and their interface accordingly specified) to take advantage of the same fundamental {QUAC} behavior to enable low-cost, high-throughput true random number generation {(which we describe next in Section~\ref{sec:quac-trng})} as intended behavior.}

\section{\XT}
\label{sec:quac-trng}
{\XT generates true random numbers at high-throughput by {repeatedly} performing \X.}

\subsection{Generating Random Output From \X}
\label{sec:quac-generatingrandomoutput}

\begin{figure}[!ht]
\centering
\includegraphics[width=0.49\textwidth]{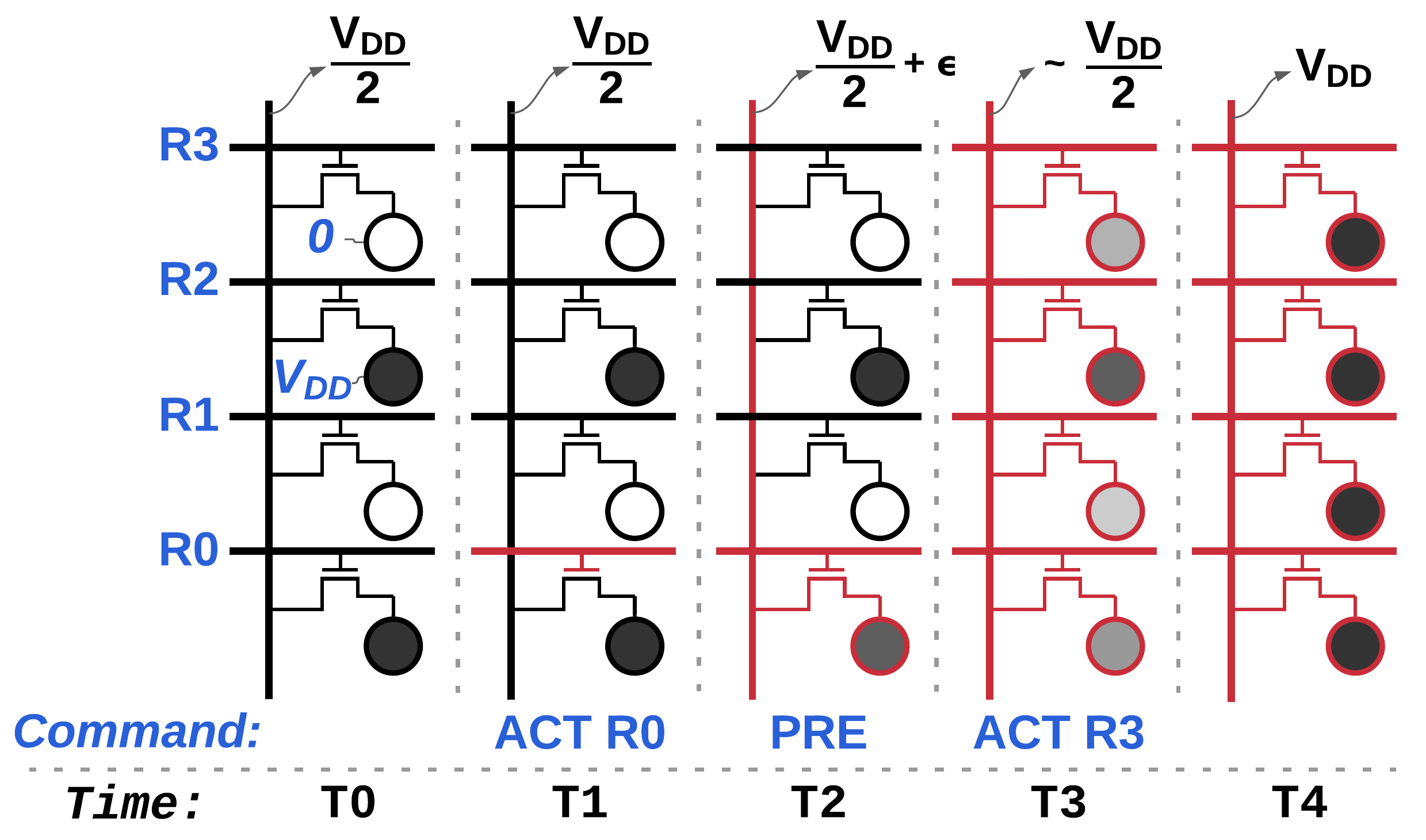}
\caption{{Timeline of c}hanges in a DRAM bitline's state in a DRAM segment during {a} \X{} {operation}. Dashed {vertical} lines represent a state transition.}
\vspace{1mm}
 \label{fig:fra}
\end{figure}

Fig{ure}~\ref{fig:fra} 
depicts {how a \X operation generates a random output when the cells {in} rows R0 and R2 are initially {\emph{charged} ($V_{DD}$)}, and the cells {in} \om{rows} R1 and R3 are initially {\emph{discharged} ($0$)} in a DRAM segment.}

At {T0}, the bitline is precharged {($V_{DD}/2$)}. At {T1}, we {enable wordline} \textit{R0} by {quickly} issuing an \textit{ACT} {command} {to R0}. We interrupt the \textit{ACT} command by issuing a \textit{PRE} command at {T2}. {Meanwhile, the cell on \textit{R0} shares a portion of its charge with the bitline, reducing its voltage level ($<V_{DD}$).} Before the \textit{PRE} command closes the row and precharges {the bitline}, we issue another \textit{ACT} command to \textit{R3} at {T3}. The last \textit{ACT} command interrupts the \textit{PRE} command and enables wordlines R1, R2, and R3 simultaneously{, in addition to the already enabled R0}. Since \X opens four rows, all four cells on a DRAM bitline contribute to the bitline voltage. {Following \X, {at T4,} the bitline end{s} up with {a} voltage level below reliable sensing margins{. Thus, it is} sampled as {a} random value by the sense amplifier{;} in Figure~\ref{fig:fra}, the {single depicted} bitline is randomly sampled as $V_{DD}$.}

To explain \X{}'s {true random number generation} behavior,
{we hypothesize that \X{} produces random values in sense amplifiers by forcing each sense amplifier to attempt to amplify a differential voltage that is well below its reliable sensing margin (i.e., there is approximately no voltage difference between {the sense amplifier's} two terminals). Under these conditions, the sense amplifier fail{s} to reliably develop and non-deterministically settle{s} to either logical high or low based on thermal noise~\cite{bhargava2015robust}{.}\footnote{{We do {\emph{not}} observe this behavior in \emph{every} DRAM bitline within a DRAM segment. We attribute this to the effects of process variation across different components in the DRAM array, e.g., the capacitance of DRAM bitlines, the offset of differential sense amplifiers and the capacitance of DRAM cells.}} To achieve this, {we} initialize the four rows that will undergo QUAC with data patterns that ensure opposite charge values in DRAM cells along the same bitline. {W}hen charge sharing occurs amongst the four cells following a \X{} operation, the bitline remains close to the quiescent bitline voltage of {$V_{DD}/2$}. Therefore, any data pattern that programs the four cells with conflicting charge values will suffice.}\footnote{{To {analyze \X's} data pattern {dependency}, we {exhaustively test \X with {16} data patterns\om{,} as we describe in Section~\ref{sec:quac-randomness}.}}} 

\subsection{{Mechanism}}
\label{sec:quac-trng-trng}
{\XT} {leverages the random values {in} the sense amplifiers generated by \X operations} as its source of entropy. \XT{} first performs a \X{} operation on a {\emph{high-entropy DRAM segment}}\footnote{{A {high-entropy segment is a DRAM} segment where \X{} operations generate {\emph{many}} random values {(i.e., with 1000s of bits of entropy)} in the sense amplifiers, identified through a one-time characterization effort\om{,} as described in Section~\ref{sec:characterization-entropy}}{.}} and generates random values {in} the sense amplifiers. \XT{} {then} uses the SHA-256 cryptographic hash{~\cite{fips2012180}} function to post-process the {random values {in} the sense amplifiers} to generate {high-quality true} random numbers. 

\begin{figure}[!h]
\centering
\includegraphics[width=0.5\textwidth]{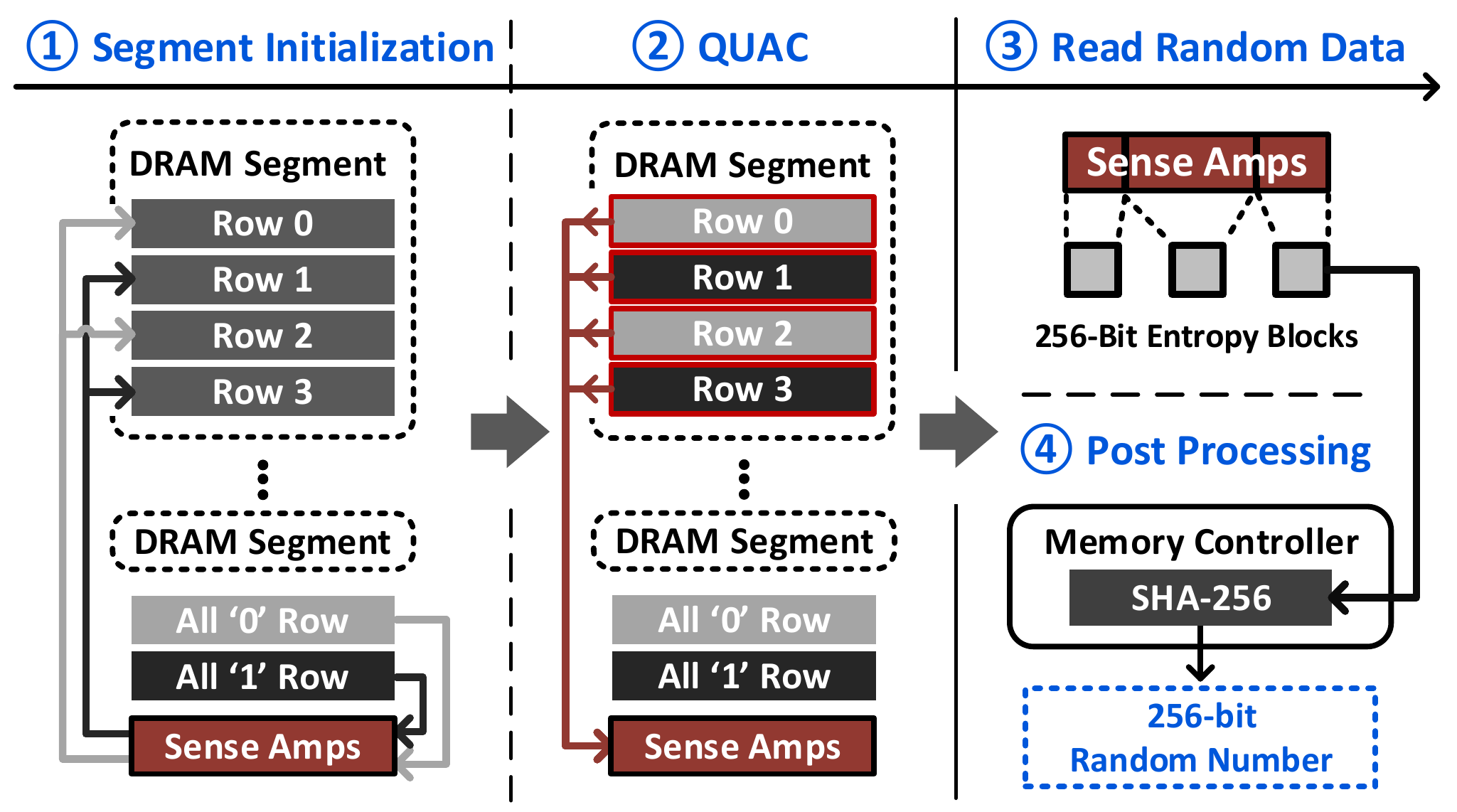}
\vspace{-5mm}
\caption{QUAC-TRNG mechanism.
}
\label{fig:mec-exp}
\end{figure}

{{Figure~\ref{fig:mec-exp} depicts a DRAM subarray's logical organization when used} for \XT and the three-step procedure of generating a 256-bit random number with QUAC-TRNG.}
{QUAC-TRNG {reserves} six rows in a DRAM subarray {to {ensure} that no other system component {can} access the reserved {rows}}. Four of these rows form a segment that is used to perform QUAC. Two of them store {\emph{all-0s}} and {\emph{all-1s}} for initializing the segment with low latency.}

To generate a 256-bit random number, \XT{} first selects a high-entropy DRAM segment and initializes the segment by performing four in-DRAM copy operations{~\cite{seshadri2013rowclone,fei2019computedram}} {from} the two reserved rows {to each row} in the segment {\hcircled{1}}. Second {\hcircled{2}}, it performs a QUAC operation on the segment to generate random data in {the} sense amplifiers. Third {\hcircled{3}}, the memory controller reads a block of bits from the {sense amplifiers with a total amount of 256 bits of Shannon entropy {(Section~\ref{sec:experimental-methodology})}.}
Finally \hcircled{4}, {the memory controller post-processes this block using {the} SHA-256 hash function to generate a 256-bit random number with improved quality of randomness.}

\section{Real {DRAM} Chip Characterization}
\label{sec:evaluation}
\subsection{Randomness in \X Operations}
\label{sec:quac-randomness}

{We} experimentally stud{y} the entropy characteristics of \X{} operations {across} different {data patterns and} DRAM segments {in real DRAM chips}.
\subsubsection{Experimental Methodology}
\label{sec:experimental-methodology}

To characterize the {entropy in random values resulting from} \X operations{,} we {conduct} experiments on {\nochips DRAM chips that come from \nodimms off-the-shelf DDR4 modules (see Appendix A, Table~\ref{table:ddr4_table})}.

\begin{figure}[!ht]
  \centering
  \includegraphics[width=0.48\textwidth]{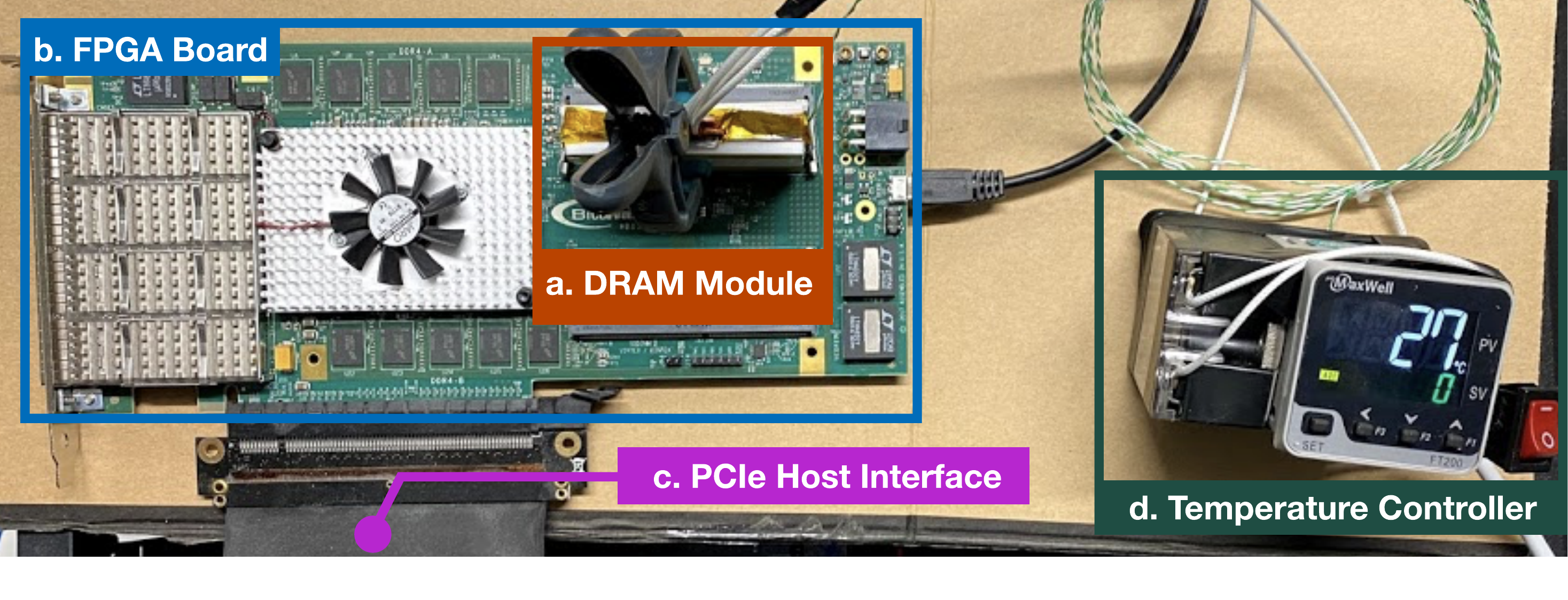}
  \vspace{-6mm}
  \caption{DDR4 SoftMC experimental setup.}
  \label{fig:softmc-setup}
\end{figure}

\noindent
{\textbf{Infrastructure.} We use a modified version of SoftMC~\cite{SoftMC} {that} enables precise control over DDR4 command {timings}{, also used in}~\cite{kim2020revisiting, frigo2020trrespass}. We test DDR4 modules (Figure~\ref{fig:softmc-setup}-{a}) by issuing DDR4 command sequences that we send to the FPGA board (Figure~\ref{fig:softmc-setup}-{b}) from the host machine through the PCIe interface (Figure~\ref{fig:softmc-setup}-c). 
During our experiments, we control the temperature of DRAM chips on both sides of the {module}. To do so, we vertically connect the {module} to the FPGA board and heat the module as needed from both sides using rubber heaters (Figure~\ref{fig:softmc-setup}-{a}). To control the heaters, we use a temperature controller {(Figure~\ref{fig:softmc-setup}-d)} that performs a closed-loop PID control{, which keeps} the temperature constant at $\pm$\SI{0.1}{\celsius} of the desired temperature level {(\SI{50}{\celsius} by default)}.}

Algorithm~\ref{alg:testing_quac} {describes} the test {procedure} we use to {extract true random numbers using QUAC operations}.
{Algorithm~\ref{alg:testing_quac} consists of three steps: {step} \One{}~initializes} the DRAM segment with a data pattern {(\emph{Line 2}), {step} \Two{}~performs} a \X operation on the DRAM segment {(\emph{Lines 3-7}), and {step} \Three{}~reads back the random values in the row buffer (\emph{Lines 9-10}).}
{To \emph{simultaneously} enable all four rows in a segment, {we} activate the first and the fourth rows in the segment (e.g., $Row_{0}$ and $Row_{3}$) with two greatly violated timing parameters, \tras{}, and \trp{}. First, {we} issue the \emph{PRE} command (\emph{Line~5}) earlier than the time delay (\tras{}) needed for charge restoration to complete.
Second, {we} issue the second activation (\emph{Line~7}) earlier than the time delay (\trp{}), needed for bitlines to settle at $V_{dd}/2$. We obey the DRAM timing parameters while reading from every sense amplifier in the DRAM segment.\footnote{{We repeat Algorithm~\ref{alg:testing_quac} for every DRAM segment in a DRAM bank in all DRAM modules.}}}

\begin{algorithm}[tbh]\footnotesize
        \SetAlgoNlRelativeSize{1.0}
        \DontPrintSemicolon
        \caption{Testing for \X's randomness}
        \label{alg:testing_quac} 

        \textbf{DRAM\_QUAC\_randomness\_testing($data\_pattern$, $DRAM\_segment$, {$DRAM\_bank$}):} \par 
        ~~~~write $data\_pattern$ into all rows in $DRAM\_segment$ \par 
        ~~~~$activate(DRAM\_segment:Row\_0)$ \par
        ~~~~$wait(2.5ns)$ {// violate \tras{}}\par
        ~~~~{$precharge(DRAM\_bank)$} \par
        ~~~~$wait(2.5ns)$ {// violate \trp{}}\par
        ~~~~$activate(DRAM\_segment:Row\_3)$ \par
        ~~~~$wait(tRCD)$ \par
        ~~~~\textbf{foreach} $SA$ in $DRAM\_segment$: ~~// read each sense amplifier\par 
        ~~~~~~~~ record the value on the $SA$ \par
    \end{algorithm}

\noindent \textbf{Shannon Entropy. }
Shannon entropy~\cite{shannon1948mathematical} {quantifies} the amount of information present in a signal. 
We use Shannon entropy as a measure {of} the randomness {in} DRAM {sense amplifiers} following \X operations. 
We calculate a sense amplifier's Shannon entropy as in {Equation~\ref{equ:background-shannon-entropy}}, where $p(x_1)$ is the probability of observing a logical-0 value and $p(x_2)$ is the probability of observing a logical-1 {value in} the sense amplifier following \X operations.
{The total Shannon entropy \om{(i.e., entropy)} of a bitstream can be interpreted as the \emph{effective} number of random bits within the bitstream.}
\vspace{-1mm}
\begin{equation}
    H(x) = -\sum_{i=1}^2 p(x_i) \log_2 p(x_i)
    \label{equ:background-shannon-entropy}
\end{equation}
\subsubsection{{Methodology to Measure} Entropy in \X Operations}
\label{sec:characterization-entropy}

{We measure the entropy of the random bitstreams generated {in} individual sense amplifiers by performing \X operations. {We} repeatedly perform \X (as shown in Algorithm~\ref{alg:testing_quac}) 1000 times {and measure the entropy {of} each sense-amplifier by evaluating Equation~\ref{equ:background-shannon-entropy} for the 1000-bit bitstream {produced by} each sense amplifier. We repeat this analysis} on \emph{8K} different DRAM segments (32K DRAM rows) {using} \emph{16} different data patterns{.}}
{We refer to the entropy of the bitstreams obtained from a sense amplifier connected to a bitline in a DRAM segment as that \emph{bitline's entropy}}. 

\subsubsection{Data Pattern Dependence}
\label{sec:characterization-dpd}

\begin{figure}[!ht]
  \centering
  \includegraphics[width=0.49\textwidth]{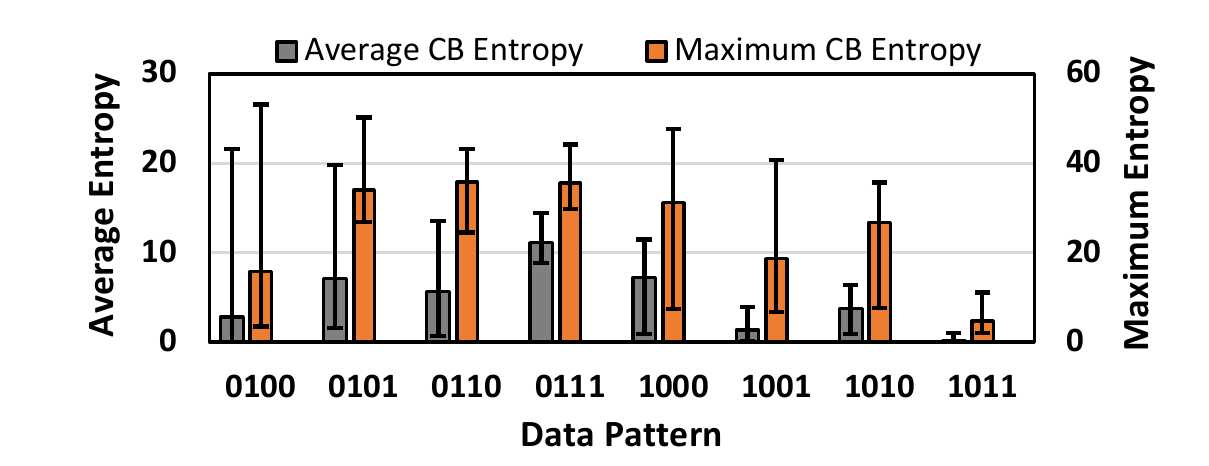}
  \caption{Average {({grey bars}, left Y-axis)} and maximum {({orange bars}, right Y-axis)} DRAM cache block entropies for different data patterns {across \nodimms modules. The error bars show the {range} of the average and the maximum DRAM cache block entropy across all modules.}}
  \label{fig:avg-ent}
\end{figure}

{We analyze how the data patterns used in initializing DRAM segments affect the result of \X operations.} We calculate the entropy for {each} cache block ({i.e.,} 512 bitlines) in a DRAM module by aggregating the entropy of all bitlines in {the} cache block. {We define two metrics \One \emph{average cache block entropy}, and \Two \emph{maximum cache block entropy}.\footnote{{The theoretical maximum entropy for a single cache block is 512 bits because each cache block is 512 bits (i.e., 64 bytes) wide.}} We calculate the \emph{average cache block entropy} as the average entropy across all cache blocks in a DRAM module. The \emph{maximum cache block entropy} is the entropy of the cache block with the highest entropy in a DRAM module. Figure~\ref{fig:avg-ent} shows the average values of each of these metrics across all \nodimms modules we test. The error bars show the {range} (i.e., minimum and maximum) of the values across all modules.}
A {larger} entropy indicates more random behavior {in} DRAM sense amplifiers. {We omit the data patterns that {result in} insufficient entropy {{in} sense amplifiers following QUAC operations}.}

We make {three} observations {from Figure~\ref{fig:avg-ent}}. 
First, the average entropy {varies across different data patterns.} {The average cache block entropy is the highest at {11.07} bits for the data pattern ``0111'' whereas it is the lowest at 0.17 bits for data pattern ``1011''.} 
Second, we observe that the ``0111'' { and ``1000''} data pattern{s} {lead to the highest entropy} {on average} in all DRAM modules we test.
This indicates that random{ness increases}
when the first row \X activates {($Row_0$)} is initialized with {the inverted value of all other three rows (e.g.,} all-zeros {in $Row_0$} and {all-ones in} the other three rows{).}
{This is because} the {cells in the} first row
have more time to share their charge with the bitlines {as they are} {activated earlier than {the} other three rows}. We hypothesize
the bitline voltage is more likely to end up at a metastable level if all three {later-activated} rows simultaneously try to pull the bitline voltage {in the opposite direction of the row {that is activated first in QUAC operations}.}
{Third, w}e observe that cache block entropy in \X operations can reach up to {53.0} bits {with the ``0100'' data pattern. We hypothesize that this is a result of {a combination of} design-induced variation \cite{lee2017design} and manufacturing process variation across DRAM segments. For example, variation in DRAM cell capacitance across DRAM segments may result in some DRAM segments to favor a certain data pattern (e.g., ``0100''), i.e., performing \X{} on this segment keeps the bitline voltage below reliable sensing thresholds when the rows are initialized with {that} data pattern.}

\subsubsection{Spatial Distribution of Entropy}
\label{sec:characterization-spatial}
\begin{figure}[!ht]
  \centering
  \includegraphics[width=0.49\textwidth]{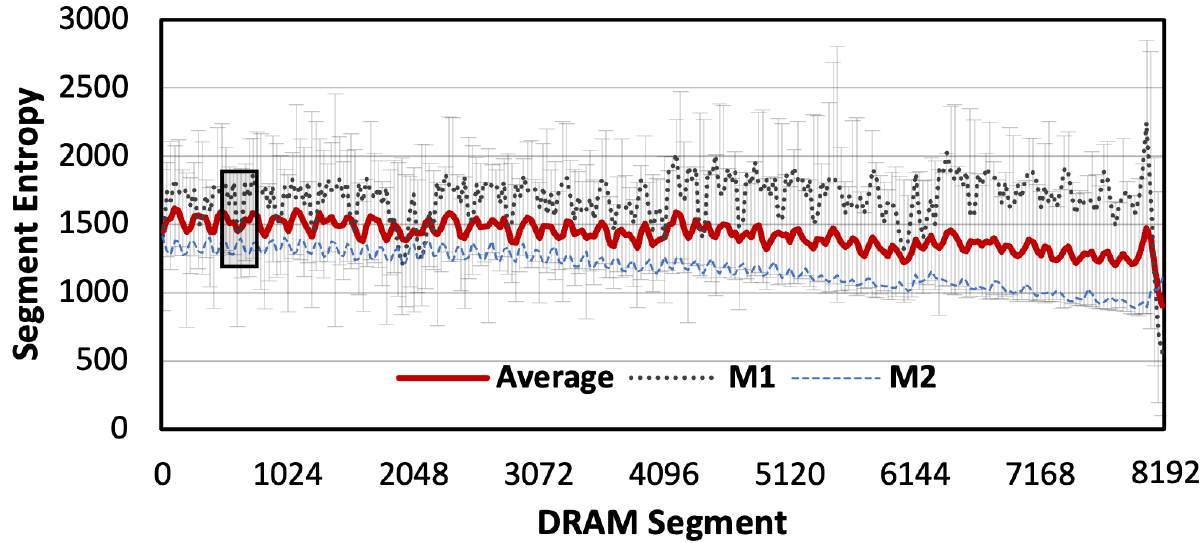}
  \caption{{Average DRAM segment entropy across {\nodimms modules (}\nochips chips{)}.} The X-axis plots the DRAM segments and the Y-axis shows the segment entrop{y}. {We plot the segment entropy of two {specific} modules {(M1 \& M2)} using black {(dotted)} {and blue {(dashed)}} lines.}}
  \label{fig:seg-ent}
\end{figure}

We {study} the spatial distribution of entropy in \X operations {across segments in a DRAM bank.}
We calculate {a segment's entropy} as the sum of all bitline entropies in a DRAM segment. Figure~\ref{fig:seg-ent} {depicts how a segment's entropy (y-axis) varies across} 8K DRAM segment{s in a DRAM {bank} (x-axis) across \nochips DRAM chips, initialized with the data pattern that yields the largest {average} entropy (``0111'').} {There are three curves in {Figure~\ref{fig:seg-ent}.}} {The red curve shows the average {segment} entropy across all chips, with the error bars showing the maximum and minimum entropy values observed for {any} DRAM segment.\footnote{{The theoretical maximum entropy of a single segment is 64K bits because there are 64K bitlines in each DRAM segment.}} Black {(dotted)} and blue {(dashed)} curves provide {representative samples of} two main entropy variation trends ({M}1 and {M}2, respectively{, depicting two selected DRAM modules}) we observe across all chips.}

We make three observations from Figure~\ref{fig:seg-ent}. First, the DRAM segment entropy behavior is different across {modules}. {For example,} the 640\textsuperscript{th} segment (middle of the highlighted area on the figure) {exhibits significantly lower entropy compared to nearby segments (i.e., {leads to} a local minimum)}
in module {{M}1}, but it {exhibits a significantly higher entropy compared to its neighboring segments (i.e., {leads to} a local maximum)}
in module {{M}2}.
{Assuming the two modules{'} circuit designs are identical (since both modules are from the same manufacturer), we {can potentially}} attribute this {difference between modules} to {systematic process variation~\cite{mehrotra2001modeling}} {and{/or}} {post-manufacturing row repair, where erroneous DRAM rows are remapped on a per-chip basis after manufacturing to improve yield{~\cite{kim2014flipping, smith1981laser, horiguchi1997redundancy, keeth2001dram, itoh2013vlsi, liu2013experimental, kang2014co, seshadri2015gather, khan2016parbor, khan2017detecting, lee2017design, tatar2018defeating, barenghi2018software, cojocar2020rowhammer,  patel2020beer}}}. 
Second, we observe that {the overall} segment entropy distribution follows a wave-like pattern{.} 
The segment entropy peaks and descends repeatedly as segment id (x-axis) increases {(i.e., as DRAM row addresses increase) in the same DRAM bank}. 
{We hypothesize that this spatial pattern results from either the effects of systematic process variation or the structure of the local DRAM array.}
{For example,} {a} segment{'s} entropy could be related to the segment's distance from the sense amplifiers.
Third, a majority of modules experience a significant increase in segment entropy towards the 8000\textsuperscript{th} segment, followed by a drop in segment entropy towards the end (i.e., 8192\textsuperscript{nd} segment) of the DRAM {bank}.
{{{T}his could {potentially} be explained by {systematic process variation or} the} {micro{-}}architectural characteristics of the DRAM {bank}. For example, the subarrays at the end of the {bank} might be differently sized than the rest of the subarrays, placing some segments further away from the sense amplifiers.}

\begin{figure}[!ht]
  \centering
  \includegraphics[width=0.49\textwidth]{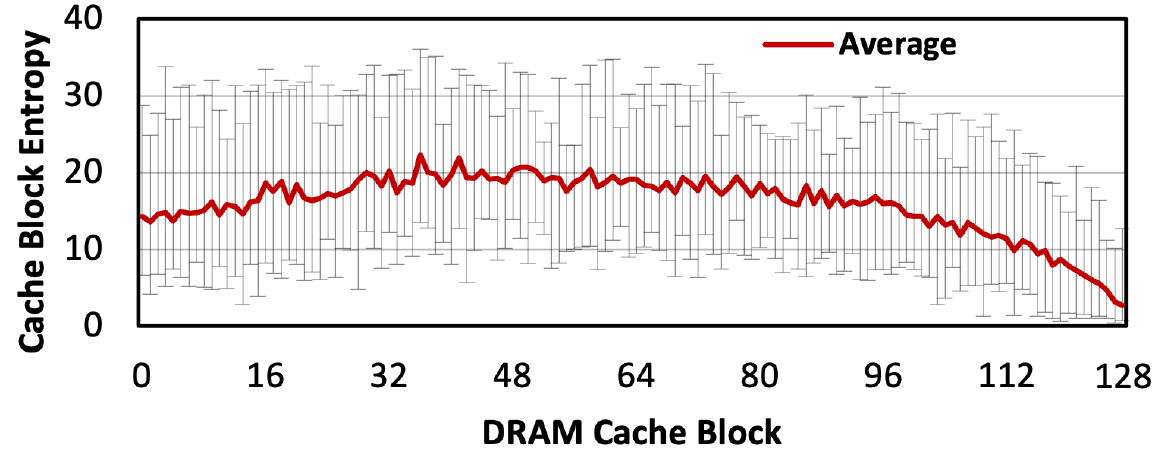}
  \caption{{{Average entropy of each cache block in the highest-entropy segment in all modules. The error bars show the {range} of the values across all modules}}.}
  \label{fig:cb-ent}
\end{figure}

{We calculate a {\emph{cache block's entropy (cache block entropy)}} as the sum of the entropy of all bitlines in that cache block. We use the highest average-entropy data pattern (``0111'') to initialize DRAM segments and find each {cache block's entropy} in the highest-entropy DRAM segment in each DRAM module. Figure~\ref{fig:cb-ent} plots the average value of each {cache block's entropy}} in the highest-entropy DRAM segment, and the error bars show the {range} (i.e., minimum and maximum) of the values across all 17 modules. We observe that the cache block entropy {peaks around the middle of the DRAM segment and} deteriorates towards the end of the DRAM segment. This indicates that the bitlines in the higher-{numbered} cache blocks are {less random than} the bitlines in the lower- {or middle-numbered} cache blocks.

We conclude from our analysis that the entropy {provided by} \X operations is distributed non-uniformly across DRAM {segments and DRAM cache blocks}. We hypothesize that the entropy distribution could be related to the {micro-}architectural characteristics of DRAM {banks} (e.g., distance {of segments from} the sense amplifiers), {systematic variation in manufacturing processes~\cite{mehrotra2001modeling}, or post-manufacturing row-repair}.

\subsection{True Random Bitlines in \X Operations}
\label{sec:nist-sts}
We conduct {a} SoftMC experiment to demonstrate that \X{} operations, when performed repeatedly, generate random bitstreams {in} DRAM sense amplifiers. The SoftMC experiment works in three steps: \One initializes the DRAM segment with a data pattern, \Two performs a \X operation on the DRAM segment to generate random values {in} the sense amplifiers, \Three reads out the DRAM segment. We collect one bit {from} each sense amplifier in the DRAM segment with each iteration of our experiment. We iterate one million times to collect 1 Mb bitstreams from every sense amplifier in the DRAM segment. Our {entropy analysis} show{s} that the values produced by \X operations on {all} sense amplifiers are biased towards a binary (logic-0 or logic-1) value {(i.e., more likely to produce either one of the binary values)}. {We use post-processing methods (Von Neumann Corrector~\cite{neumann1951various} {and} SHA-256~\cite{fips2012180}) to improve the quality of random bitstreams generated by QUAC operations.} 

We apply the Von Neumann Corrector {(VNC)}~{\cite{neumann1951various}} to all bitstreams to remove bias and improve the quality of the random number sequence. The {VNC} first splits all bits into groups of two bits. Then it applies one of the three transformations: \One removes the group if both of the bits have the same value, \Two removes the group and inserts a logic-1 if the first bit in the group is logic-0 and the second one is logic-1 (i.e., the generator transitions from logic-0 to logic-1), or \Three removes the group and inserts a logic-0 otherwise. E.g., the bitstream ``0010'' after post-processing using the {VNC} becomes ``0''.

We use the NIST Statistical Test Suite {(STS)}~\cite{rukhin2001statistical} to validate the {randomness of the }output of our TRNG. NIST STS formulates several statistical tests to test a specific \emph{null hypothesis}, H\textsubscript{0}, which states that the number sequence under test is \emph{random}. The suite outputs a \emph{p-value} for all of the statistical tests that it runs on the random number sequence. We say that H\textsubscript{0} holds for a statistical test if it outputs a p-value greater than a chosen \emph{level of significance} denoted as $\alpha$. {That is}, if the p-value of a test is greater than $\alpha$, then the number sequence is random according to that test. We choose $\alpha$ as 0.001 based on the suggested \emph{level of significance} range ([0.01, 0.001]) in the NIST STS specification~\cite{rukhin2001statistical}.

We collect bitstreams from every sense amplifier (64K in one DRAM segment) in a DRAM segment following \X operations. We test 8K DRAM segments in every DRAM module. {We observe that 1 Mbit bitstreams collected from {22} sense amplifiers can pass all NIST STS tests.}

{Table~\ref{table:nist-sts} presents the {average} p-values for {the} NIST STS test {results} on two types of bitstreams {that pass all 15 tests}: \One the output of {the} Von Neumann Corrector (``VNC'') and \Two the output of the post-processing {step} we describe in Section~\ref{sec:quac-trng-trng} (``SHA-256''). {We conclude} that \X generates number sequences that are indistinguishable from {true} random number sequences. We discuss the randomness of post-processed results (SHA-256 column) in Section~\ref{sec:quac-trng-output-eval}. }

\begin{table}[h]
    \vspace{0.5em} 
    \renewcommand{\arraystretch}{0.85}
    \caption{NIST STS {Randomness Test} Results}
    \centering
    \footnotesize
    \vspace{-0.5em} 
    \begin{tabular}{  l r r }
    \multirow{2}{*}{\textbf{NIST STS Test}} & \textbf{VNC$^*$} & \textbf{SHA-256} \\
                                            & \textbf{\footnotesize(p-{v}alue)} & \textbf{\footnotesize(p-{v}alue)}\\
    \hline
    monobit & {0.430} & {0.500} \\
    frequency\_within\_block & {0.408} & {0.528}\\
    runs & {0.335} & {0.558}\\
    longest\_run\_ones\_in\_a\_block & {0.564} & {0.533}\\
    binary\_matrix\_rank & {0.554} & {0.548}\\
    dft & {0.538} & {0.364}\\
    non\_overlapping\_template\_matching & $>$0.999 & {0.488}\\
    overlapping\_template\_matching & {0.513} & {0.410}\\
    maurers\_universal & {0.493} & {0.387}\\
    linear\_complexity & {0.483} & {0.559}\\
    serial & 0.3{55} & {0.510}\\
    approximate\_entropy & {0.448} & {0.539}\\
    cumulative\_sums & {0.356} & {0.381}\\
    random\_excursion & {0.164} & {0.466}\\
    random\_excursion\_variant & {0.116} & {0.510}\\
    \hline
    \multicolumn{3}{c}{$^*$VNC: Von Neumann Corrector}
    \end{tabular}
    \label{table:nist-sts}
\end{table}

\section{\XT Evaluation}
\label{sec:quac-trng-eval}

{{W}e ev{a}luate \XT using real {DRAM} chip experiments and simulation studies to show that {\XT} {\One} produces high-quality random bitstreams, and {\Two} outperforms prior DRAM-based TRNG proposals.}

\subsection{{\XT Output Quality}}
\label{sec:quac-trng-output-eval}
{To demonstrate that \XT produces high-quality bitstreams of random values, we experimentally extract {nine} bitstreams from {three DDR4 modules (24 DRAM chips)}.{\footnote{We test a total {of} {nine} bitstreams, each sized \SI{1}{\giga\bit}, obtained from three DRAM modules to demonstrate that QUAC-TRNG can produce statistically uncorrelated streams of random numbers {while} maintain{ing} a reasonable testing time.}} Our results show that the bitstreams pass {all of} the NIST STS {tests}.}

{We extract a single bitstream using} five steps: {we} \One initialize the DRAM segment with the \emph{highest-entropy} data pattern {(``0111'')}{,} \Two perform a \X operation on the DRAM segment{,} \Three read out the DRAM segment{,} \Four split the DRAM segment into blocks {that each ha{ve}} 256 bits of entropy {based on our characterization of cache block entropy in Section~\ref{sec:characterization-entropy}, and \Five} input the 256-bit entropy blocks to the SHA-256 hash function to obtain 256-bit random number{s}. 

{We} {partition \SI{1}{\giga\bit} bitstreams obtained from each high{est-}entropy DRAM segment into \SI{1}{\mega\bit} random number sequences and} test
1024 number sequences {per} {DRAM segment} using NIST STS. 
{We find that} 
$99.{28}\%$ of the sequences pass all NIST STS tests. {This pass rate is larger than the acceptable rate\footnote{Based on the formula $(1-\alpha)\pm3\sqrt{\alpha(1-\alpha)/k}$, where k is the sequence population (${1024}$) and $\alpha$ is the significance level ($0.00{5}$)} ($98.84\%$) that NIST specifies{~\cite{rukhin2001statistical}}.}

{Table~\ref{table:nist-sts}}{, column ``SHA-256''} shows the average p-value for
{each test.}
We conclude that \XT generates {high-quality} uncorrelated, random bitstreams.

\subsection{\XT Throughput}
\label{subsec:empirical_tput}

We analytically {model} \XT's throughput {for a module} {in terms of \One} the number of input blocks with 256 bits of entropy in the high{est-}entropy segment {($SIB$: SHA Input Blocks)} {and \Two} the overall latency of one \X operation {($L$). \XT generates $256 \times SIB$ random bits per DRAM bank in $L$ \SI{}{\nano\second}, resulting in a throughput of $(256 \times SIB) / (L \times 10^{-9})$ bits per second.}
$SIB$ is calculated directly from the {entropy of the} high{est}-entropy segment as $\floor{segment\_entropy/256}$. {We calculate $L$ by tightly scheduling the DRAM commands required to (i) initialize four DRAM rows with data patterns, (ii) perform QUAC, and (iii) read random values {from} the sense amplifiers {in}to the memory controller.}

{\XT's latency {($L$)}} is dominated by the time it takes to initialize four DRAM rows in a DRAM segment. We apply two optimizations to {amortize} the initialization overhead
and increase the peak throughput of \XT. First, we {concurrently execute \X{} {operations} {across} multiple banks by exploiting bank-level parallelism. In particular, for DDR4, we interleave across bank groups due to {DDR4's} short ACT-to-ACT {($tRRD\_S$)} timing constraint.}
Second, we use \emph{in-DRAM copy} operations to initialize DRAM segments {at} row granularity {by adopting ComputeDRAM's~\cite{fei2019computedram} {RowClone-based~\cite{seshadri2013rowclone}} in-DRAM copy procedure {in} our DDR4 modules.} Using in-DRAM copy, we significantly reduce {the} DRAM segment {initialization latency}. 

\begin{figure}[!ht]
\centering
\vspace{1mm}
\includegraphics[width=0.49\textwidth]{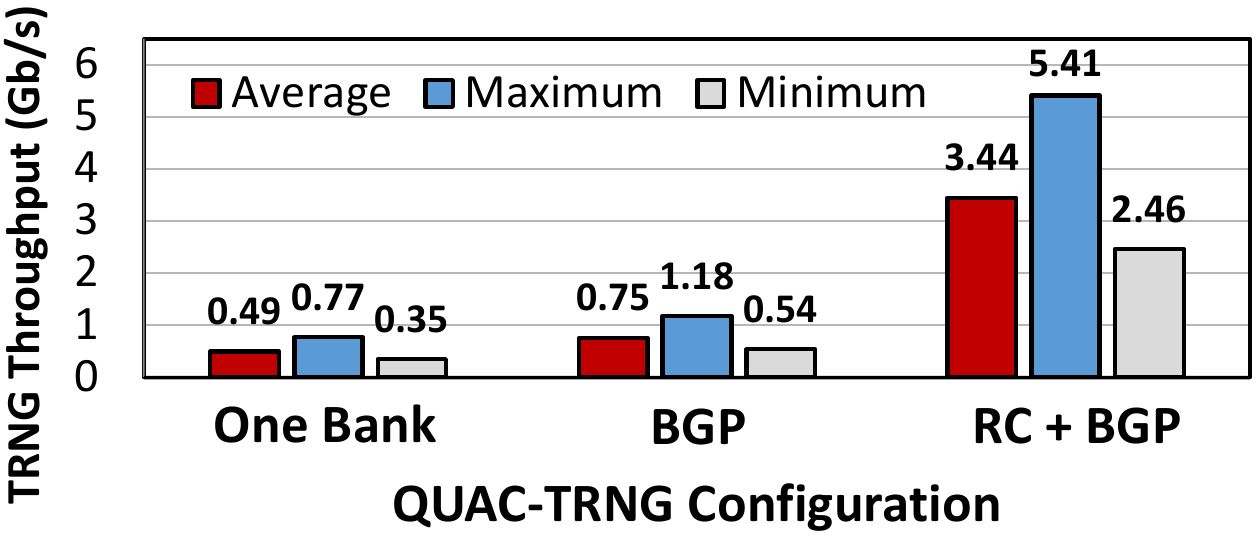}
\caption{\XT's random number generation throughput {(per DRAM channel)} {under three (One Bank, BGP, RC + BGP) configurations}{.}}
\label{fig:rand-throughput}
\end{figure}

Figure~\ref{fig:rand-throughput} {shows} {\XT's} random number throughput under three configurations: \One \emph{One Bank}, where we use a single DRAM bank to generate random numbers, \Two \emph{BGP} {(Bank Group Parallelism)}, where we use four bank{s from different bank} groups and overlap DRAM command latencies to fully utilize the available DRAM bandwidth, {and} \Three \emph{RC} {(RowClone)} + \emph{BGP}, where we initialize DRAM segments using in-DRAM copy to alleviate the overheads of segment initialization and use four {banks from different} bank groups. {We plot the average, maximum, and minimum TRNG throughput \XT{} provides across all DRAM modules.}

We observe that, on average, \emph{One Bank} achieves {0.49} Gb/s, \emph{BGP} achieves {0.75} Gb/s, and \emph{RC + BGP} achieves {3.44} Gb/s random number throughput. {The TRNG throughput of \XT{} varies across modules as the maximum segment entropy for each module varies.} We conclude that \XT{} {greatly benefits from} in-DRAM copy to achieve high {true} random number generation throughput.

\subsection{{System Performance Study}}
\label{sec:ramulator-analysis}
\begin{figure}[!ht]
  \centering
  \includegraphics[width=0.49\textwidth]{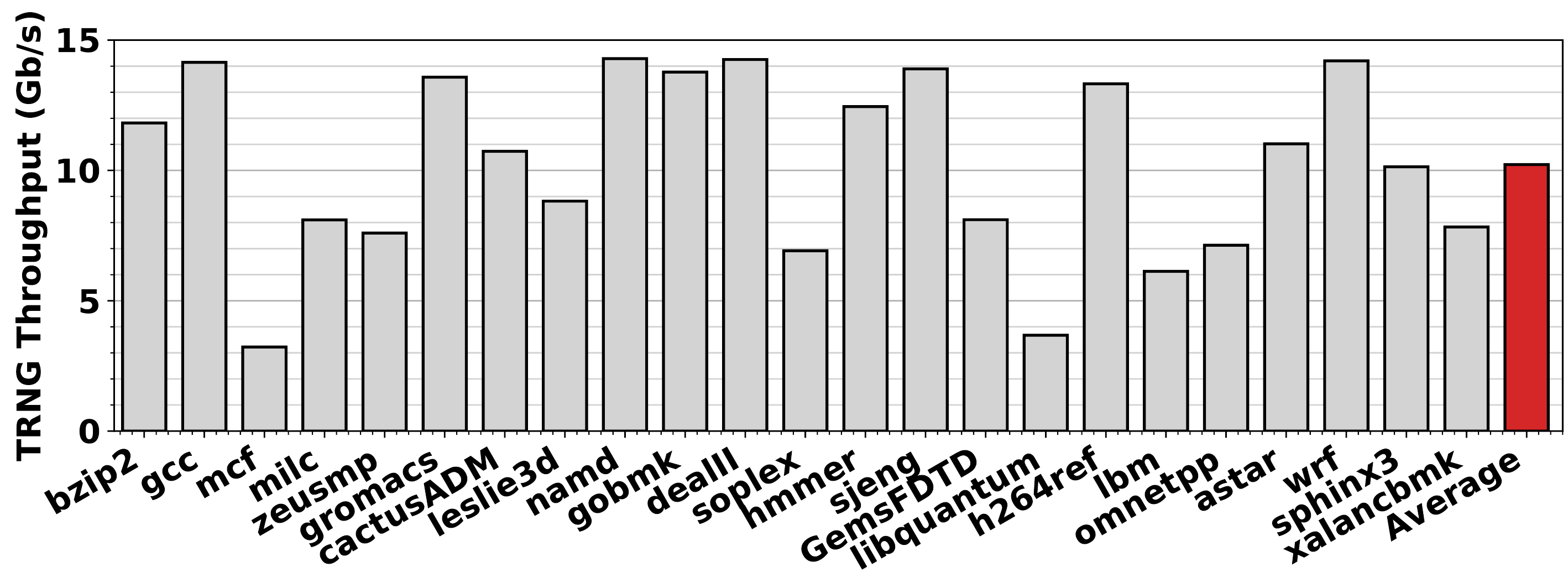}
  \caption{Available {T}RNG throughput {during idle DRAM cycles} wh{ile} running SPEC2006 workloads.}
  \label{fig:segent_bandwidth}
  \end{figure}

{To understand the maximum throughput that \XT can provide \emph{without} {reducing the \emph{total off-chip memory bandwidth} available to }concurrently-running applications,} we run an experiment using memory traces from the SPEC2006 benchmark suite. We simulate a 3.2 GHz core with four {DRAM} channels of DDR4 memory using Ramulator~\cite{kim2016ramulator,ramulatorgithub} to calculate the time each memory channel spends {idle}. We inject DDR4 commands that are issued in \XT iterations into these idle intervals. {Figure~\ref{fig:segent_bandwidth} shows the random number {generation} throughput \XT provide{s} while each SPEC2006 workload is running.\footnote{{We use four banks from different bank groups in each channel.}}} {\XT generates random numbers at 10.2 Gb/s on average with a minimum (maximum) throughput of 3.22 Gb/s (14.3 Gb/s).} {We observe that by fully utilizing the idle intervals in the memory channels, QUAC-TRNG achieve{s} on average{,} {74.13}\% of the empirical {average} throughput determined in Section~\ref{subsec:empirical_tput} {(i.e., 13.76 Gb/s for 4 DRAM channels)}.}

\subsection{Comparison With Prior Work}
\label{sec:comparison-high-throughput}

{We quantitatively {compare} high-throughput ($>100Mb/s$) DRAM-based TRNGs with \XT in this section.}
We scale {each prior work's} {T}RNG throughput and latency {according} to the simulated system {with 4 DRAM channels} described in Section~\ref{sec:ramulator-analysis}. Table~\ref{table:comparison_rngs} presents a summary of our analysis, including the \emph{low throughput {($<100 Mb/s$)}} TRNGs, which we {briefly discuss} in Section~\ref{sec:related-work}.

\begin{table}[ht]
    \caption{Summary of prior DRAM-TRNGs {vs \XT{}}}
    \vspace{-1.5mm}
    \centering
    \renewcommand\arraystretch{0.6}
    \setlength{\tabcolsep}{4pt}
    \footnotesize
    
    \hspace{1em} 
    \begin{tabular}{llrrr}
    \toprule
    \textbf{Proposal} & \textbf{Entropy} & \textbf{TRNG} & \textbf{256-bit TRNG}\\
                      & \textbf{Source} & \textbf{Throughput} & \textbf{Latency}\\
    \midrule
    \midrule
    \textbf{\XT} & Quadruple ACT & {13.76}~Gb/s & {274 ns} \\
    Talukder+{~\cite{talukder2019exploiting}} & Precharge Failure & 0.68 - {6.13}~Gb/s & {{249} ns - {201} ns}\\
    D{-}RaNGe{~\cite{kim2019drange}} & Activation Failure & 0.92 - {9.73}~Gb/s & {{260} ns - {36} ns}\\
    D-PUF{~\cite{sutar2016d}} & Retention Failure & 0.20~Mb/s & {40 s}\\
    DRNG{~\cite{eckert2017drng}} & DRAM Start-up & N/A & {700 \textmu{}s}\\
    Keller+{~\cite{keller2014dynamic}} & Retention Failure & 0.025 Mb/s & {40 s}\\
    Pyo+{~\cite{pyo2009dram}} & DRAM Cmd Schedule & 2.17 Mb/s & {112.5 \textmu{}s}\\
    \end{tabular}
    \label{table:comparison_rngs}
\end{table}

We {rigorously} compare {\XT{} to} two {state-of-the-art} works that propose high-throughput DRAM-based TRNGs~\cite{kim2019drange,talukder2019exploiting}. We calculate {both \One} the maximum {random number generation} throughput and {\Two} the minimum {latency for generating} 256-bit{ random numbers for each of the high-throughput TRNGs.}
{To do so, we} tightly schedul{e} the sequence of DDR4 commands {each TRNG} need{s} to issue{.} 

\subsubsection{D-RaNGe~\cite{kim2019drange}}
{D-RaNGe generates random numbers in DRAM by leveraging failures due to reading a cache block before the row activation latency (\trcd{}) is satisfied{~\cite{kim2019drange}}}.
We analyze the throughput of D-RaNGe under two configurations: \One \emph{D-RaNGe{-Bas{ic}}}, where we evaluate D-RaNGe as {proposed in~\cite{kim2019drange}}, and \Two \emph{D-RaNGe{-Enhanced}}, where we characterize the entropy in {\trcd{} failures} {in real DDR4 devices} to estimate the throughput of D-RaNGe combined with post-processing.

\noindent
\textbf{D-RaNGe-Bas{ic}.}
{We calculate the throughput of D-RaNGe-Bas{ic} by carefully scheduling the required DDR4 commands to induce activation latency failures and read a cache block. For our analysis, we augment D-RaNGe-Bas{ic} to exploit bank-group-level parallelism in DDR4 devices.} {D-RaNGe observes that} there are as many as four TRNG cells per cache block.
{We optimistically use the largest observed randomness (4 bits in a cache block) {in calculating D-RaNGe-Bas{ic}'s throughput}.}
{{We do not use in-DRAM copy operations to further improve D-RaNGe-Bas{ic}'s throughput} {because} D-RaNGe does not benefit from the highly parallel DRAM row initialization provided by in-DRAM copy operations. D-RaNGe only needs to initialize one DRAM cache block, which can be done efficiently using DRAM write commands.} 
{Based on these observations and assumptions, we estimate D-RaNGe-{Basic}'s maximum throughput as {916.9}~Mb/s and minimum latency for generating 256-bit random numbers as \SI{260}{\nano\second}.}

\noindent
\textbf{D-RaNGe-Enhanced.}
\label{sec:enhanced-drange}
{To calculate D-RaNGe-Enhanced's TRNG throughput, we evaluate {\nochips real DDR4 chips from \nodimms DDR4 modules} using SoftMC and find the average cache block entropy {provided by} activation latency failures.}
For each DRAM cache block in a DRAM bank, one iteration of our {SoftMC} experiment: \One initializes one DRAM row with an all-0s data pattern  (found to induce the most random behavior~\cite{kim2019drange}) and \Two accesses the DRAM row with reduced \trcd{}. We repeat this experiment \emph{1000} times and calculate {each} cache block{'s} entrop{y.} {We find the maximum cache block entropy for each DRAM module. We find the average of the maximum cache block entropy across all DRAM modules to {calculate} how many times D-RaNGe-Enhanced needs to access DRAM with reduced tRCD to gather sufficient entropy (256-bits). On average, D-RaNGe-Enhanced can harness 46.55 bits of entropy from a DRAM cache block {(out of 512 bits of theoretical maximum entropy)}.}
We {calculate} that D-RaNGe-Enhanced needs to perform {\emph{6} reduced \trcd{} accesses to generate a 256-bit random number}. 
{For a fair comparison, we apply the same post-processing (SHA-256) to D-RaNGe's output as {we do in} \XT.} 
D-RaNGe {with post-processing achieves} 
up to {9.73}~Gb/s throughput. 
{D-RaNGe{-Enhanced}'s} latency of generating a 256-bit random number is {\SI{36}{\nano\second}, including the latency of the SHA-256 hash function}. We conclude that post-processing using SHA-256 can significantly improve D-RaNGe's {T}RNG throughput as it {enables utilizing} a larger portion of the cache block for random number generation.

\subsubsection{Talukder+~\cite{talukder2019exploiting}}
Talukder et al. {propose generating random numbers in DRAM by leveraging bit failures due to activating a DRAM row \emph{before} bitlines are precharged to {V\textsubscript{DD}/2}~\cite{talukder2019exploiting}.} 
The authors use SHA-256 to post-process bitstreams that are read from DRAM. Talukder+'s mechanism {(i) induces} precharge latency failures on multiple DRAM rows, {(ii)} accumulates the random failures {in} DRAM cells{, (iii)} reads these DRAM cells{, (iv)} post-processes them using the SHA-256 hash function. {We augment their algorithm to exploit bank-group-level parallelism in DDR4 devices. We use in-DRAM copy to initialize rows before inducing precharge latency failures.} We analyze the throughput of Talukder+'s mechanism under two configurations: \One \emph{Talukder+{-Bas{ic}}}, where we estimate the throughput of the mechanism based on the authors' analysis on random cells, \Two \emph{Talukder+{-Enhanced}}, where we characterize the entropy {provided by} precharge latency failures {in real DDR4 devices} to estimate the throughput.

\noindent
\textbf{Talukder+{-Bas{ic.}}}
{We calculate Talukder+-Bas{ic}'s TRNG throughput using the results provided by the authors.} The authors report that, on average, there are 130.6 random cells in a DRAM row. To accumulate 256-bits of entropy in input {blocks of} the SHA-256 hash function, Talukder+'s mechanism needs to read 3 DRAM rows. Based on this, the throughput of Talukder+'s mechanism is 681.2 Mb/s, and the latency of generating a 256-bit random number is {\SI{249}{\nano\second}}.

\noindent
\textbf{Talukder+{-Enhanced}.}
{To calculate Talukder+{-Enhanced}'s TRNG throughput, we evaluate {\nochips real DDR4 chips from \nodimms DDR4 modules} using SoftMC and find the average DRAM row entropy {(i.e., the sum of the entropy of all bitlines in a DRAM row)} in precharge latency failures.}
{We find the \emph{maximum {row} entropy} for each DRAM module. We find the average of the {maximum {row} entropy} across all DRAM modules to {calculate} how many SHA-256 input blocks with sufficient entropy (256-bits) that Talukder+-Enhanced can extract from a {high-entropy DRAM row}. We find that, on average, Talukder+-Enhanced can harness 1023.64 bits of entropy from a {high-entropy DRAM row} {(out of 64K bits of theoretical maximum entropy)} following reduced \trp{} accesses. On average, Talukder+-Enhanced can extract {3 SHA-256 input blocks with sufficient entropy from a DRAM {row}}.}
We {calculate} Talukder+{-Enhanced}{'s throughput as 6.13~Gb/s.}
The latency of generating a 256-bit random number for the Talukder+{-Enhanced} is \SI{201}{\nano\second}. 

\begin{figure}[!ht]
\centering
\includegraphics[width=0.49\textwidth]{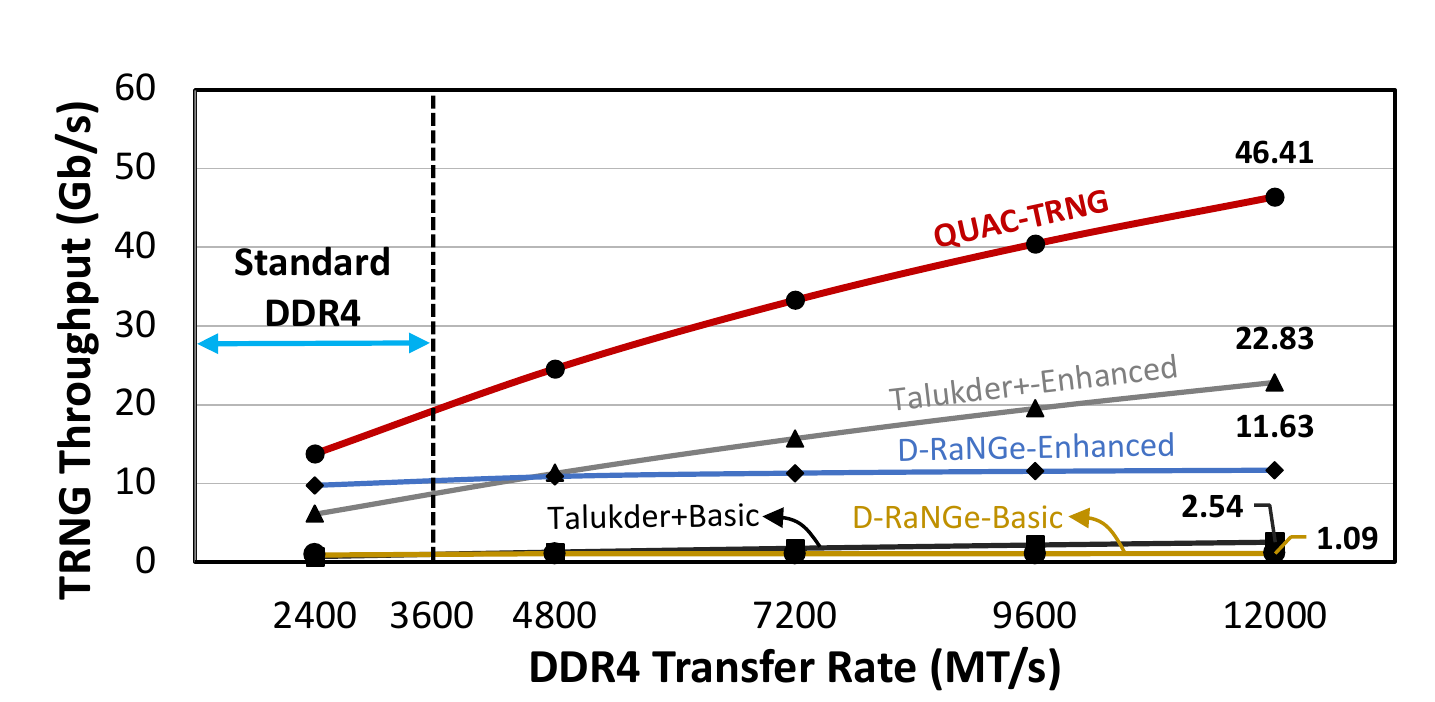}
\caption{Throughput of DRAM-based TRNGs projected on DDR4 transfer rate. We plot transfer rates beyond the DDR4 {s}tandard~\cite{jedecDDR4}.}
\label{fig:throughput-projected}
\end{figure}

Figure~\ref{fig:throughput-projected} plots the {average} throughput of Talukder+{-{Basic/Enhanced}}, D-RaNGe{-{Basic/Enhanced}}, and \XT. We project the throughput of the evaluated mechanisms {to various DDR4} data transfer rates (MT/s). 

We make two observations. {First, D-RaNGe cannot make use of the additional DRAM bandwidth because D-RaNGe needs to frequently induce activation latency failures to sustain the high throughput of random numbers. Therefore, D-RaNGe's peak throughput is bound by {DRAM} access latency and does not scale with increasing} DRAM external bandwidth. Second, Talukder+ and \XT can scale with increasing DRAM transfer rate as they are bound by the DRAM bandwidth. \XT outperforms the {basic} {(enhanced)} versions of \emph{Talukder{+}} and \emph{D-RaNGe} by \textbf{20.20$\times{}$} {(2.24$\times{}$)} and \textbf{15.08$\times{}$} {(1.41$\times{}$)}, respectively, at DDR4 2400 MT/s. At a future 12 GT/s transfer rate, \XT outperforms \emph{{enhanced}} configurations of \emph{Talukder+} and \emph{D-RaNGe} in {T}RNG throughput by \textbf{2.03$\times{}$} and \textbf{3.99$\times{}$}, respectively.

{{Although QUAC-TRNG has a higher latency than Talukder+ and D-RaNGe, this} latency for generating true random numbers can be hidden by accumulating random numbers in a buffer. Commodity systems that employ TRNGs already implement buffers to store random numbers~\cite{amd2017random}. QUAC-TRNG can fill this buffer at a significantly higher rate compared to state-of-the-art DRAM TRNGs because QUAC-TRNG achieves greater throughput.}

\section{{{Sensitivity Analysis}}}
\label{sec:discussion}
\begin{figure}[!ht]
    \centering
    \includegraphics[width=0.49\textwidth]{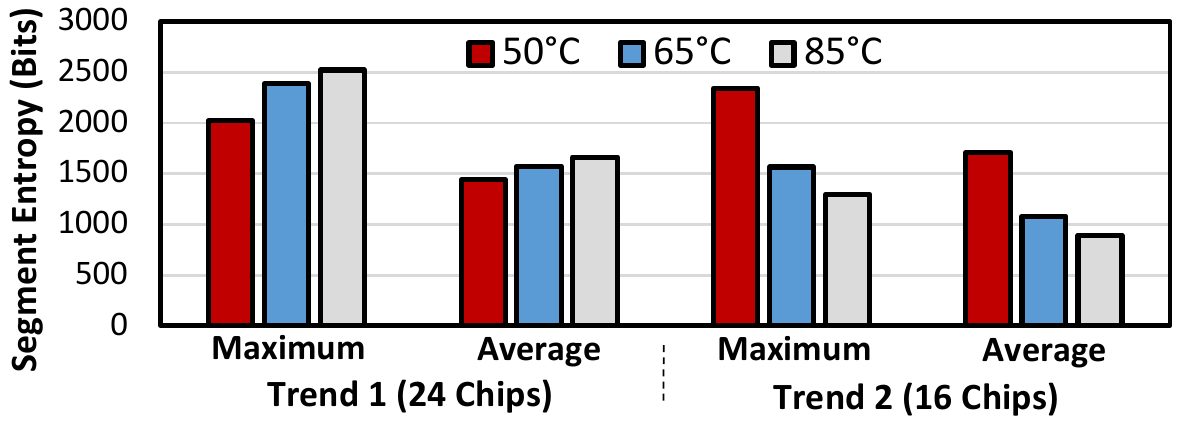}
    \caption{Maximum and average segment entropy at different temperatures.}
    \label{fig:temp-segent}
\end{figure}

\noindent
\textbf{Temperature Dependence.}
We study the effects of temperature on the entropy of \X operations {by recording} bitline entropies at {50}$^{\circ}$C, {65}$^{\circ}$C, and 85$^{\circ}$C on {40 real DRAM chips {from} 5 DRAM modules}. {We observe two trends: \emph{Trend-1}, bitline entropy increases with temperature (24 chips), and \emph{trend-2}, bitline entropy decreases with temperature (16 chips). We calculate the maximum and the average segment entropy {(sum of all bitline entropies in that segment)} independently for chips that follow \emph{trend-1} and \emph{trend-2}. Figure~\ref{fig:temp-segent} plots the maximum and average segment entropy at {50}$^{\circ}$C, {65}$^{\circ}$C, and 85$^{\circ}$C.}

We observe that the entropy in \X operations {changes} with temperature. {The maximum (average) segment entropy is 2019.6 (1442.0), 2389.8 (1569.5) and 2520.1 (1659.6) at {50}$^{\circ}$C, {65}$^{\circ}$C and 85$^{\circ}$C for DRAM chips that follow \emph{trend-1}, respectively. The maximum (average) segment entropy is 2344.2 (1710.6), 1565.8 (1083.1) and 1293.5 (892.5) at {50}$^{\circ}$C, {65}$^{\circ}$C and 85$^{\circ}$C for DRAM chips that follow \emph{trend-2}, respectively. {We conclude that a \XT{} implementation needs to account for changes in temperature while generating true random numbers, as segment entropy changes with temperature.}}

To maintain the same amount of entropy {(256-bits)} in SHA-256 input blocks {at different temperatures}, the memory controller stores a list of \emph{{column address sets}} for non-overlapping temperature ranges. {This list is initialized by identifying high-entropy DRAM segments at different temperatures during a one-time offline characterization step.}
\XT accesses an element in the list depending on {DRAM temperature (e.g., {measured} via temperature sensors~\cite{jedecDDR4})} {and retrieves a set of column addresses{, where each address} points to a contiguous range of cache blocks in the DRAM segment with 256-bits of entropy. QUAC-TRNG uses these sets} to split the data read from the high-entropy DRAM segment into SHA-256 input blocks. {In this way, QUAC-TRNG {ensures} that SHA-256 input blocks always contain 256-bits of entropy at different temperatures.}

\noindent
{\textbf{Time Dependence.}
To understand whether the quality of the random numbers that {\XT{}} generates changes over time, {we study the entropy generated by \X operations at the beginning and end of a 30-day period {using 40 chips from five modules.}}} The \emph{average segment entropy} for the highest-entropy data pattern {(``0111'', Section~\ref{sec:characterization-entropy})} does not change significantly. The difference between the average entropy of 8K segments at the beginning and at the end of the testing period is {on average (\emph{maximum, minimum}) 2.4\% (\emph{5.2\%, 0.9\%}) across five modules {(see Appendix A, Table~\ref{table:ddr4_table})}}. {We conclude that the entropy generated by \X operations is not significantly affected by time elapsed on the order of a month, so the characterized segment entropy is valid for \emph{at least} 30 days. Therefore, in the worst-case, \XT needs to re-characterize segment entropy only once a month.}

\section{{System Integration}}
\label{sec:integration}
{We discuss how \XT can be integrated into a real system. \XT{} generates random values by repeatedly (i) performing \X on the \emph{highest-entropy} (Section~\ref{sec:characterization-spatial}) DRAM segments in four banks from four {different} DRAM bank groups, and (ii) post-processing the result of \X operations using the SHA-256 hash function.}

\noindent
\textbf{Post Processing.}
\XT uses a cryptographic hash function to post-process random bitstreams produced by \X operations. We choose to evaluate \XT using SHA-256 {as the post-processing function} since SHA-256 is a secure cryptographic hash function that can be implemented efficiently in hardware {at low area and latency costs~\cite{baldanzi2020cryptographically, satoh2005ASIC, helionipcores}. This makes SHA-256 well-suited to implementation in the memory controller. We account for the costs of SHA-256 hardware in our evaluations based on values reported by recent work~\cite{baldanzi2020cryptographically}: 65 clock cycle latency {(at 5.15 GHz)}, 19.7 Gb/s throughput, and 0.001 $mm^{2}$ area at {a} 7 nm process technology {node}.}

\noindent
\textbf{\XT User Application Interface.}
{\XT generates random numbers using \X operations. {To perform \X operations, the memory controller needs to issue an ACT $\rightarrow{}$ PRE $\rightarrow{}$ ACT command sequence with reduced \tras{} and \trp{} timing parameters.} {Upon receiving a request for a random number, the memory controller checks if there is available DRAM bandwidth to perform QUAC operations and issues the command sequence with reduced timing parameters. This functionality can be implemented in a simple state machine in the memory controller's command scheduling logic.} To {eliminate delays when an application requests random numbers, the memory controller may periodically utilize available DRAM bandwidth to generate and store random numbers} in a small buffer in the memory controller, as proposed in D-RaNGe~\cite{kim2019drange}. {In this way, an application's request for random numbers can be fulfilled immediately (up to the {buffer} size).}}

In order to use \XT in a real system, the designer needs to expose an interface to user applications. {There are numerous possible ways to implement this interface, including memory- or PCIe- mapped configuration status registers, CPU co\taberk{-}processor and I/O instructions, and specialized extensions to the ISA. We leave {it to} the system designer to choose the best approach that meets the design goals for their system.}

\noindent
\textbf{Memory Overhead.}
\XT allocates a small number of DRAM rows from one bank in {four bank groups}. We allocate one DRAM segment (four rows) to {perform} \X operations {on} and two DRAM rows to initialize the DRAM segment using in-DRAM copy operations. {To fully utilize the DDR4 bandwidth, \XT{} {simultaneously} activates four segments in four bank groups (one bank in each bank group) and reads data from each bank in an interleaved manner. (Section~\ref{subsec:empirical_tput}). {Thus, w}e allocate four segments (for \X{}) and 8 DRAM rows (for bulk initialization) {across} four banks in different bank groups.} {This amounts to} 192 KB {of total reserved space}, which makes up {only} {{0.002}\% of the capacity of a{n 8} GB DDR4 module.}

\noindent
\textbf{Area Overhead.}
\XT stores \emph{4 DRAM row addresses} to point to {the starting row address{es} of the highest-entropy DRAM} segment{s} and \emph{8 DRAM row addresses} to point to the source operands for in-DRAM copy operations in four DRAM banks {from four different bank groups}. \XT also stores \emph{{11} DRAM column addresses}\footnote{{To sustain {the maximum} 5.4 Gb/s TRNG throughput {(Section~\ref{subsec:empirical_tput})} in modules where {there are} 11 {SHA-256 input blocks with 256-bits of entropy in} the highest-entropy segment.}} to indicate the non-overlapping cache block ranges that contain 256-bits of entropy each. These cache block ranges change according to system temperature (Section~\ref{sec:discussion}). We assume there are as many as 10 distinct temperature ranges in {calculating} the area overhead. In total, to store the row and column addresses, \XT uses 1{3}16 bits of storage. We model the required area for this storage using CACTI~\cite{cacti} and find that it is 0.0003 $mm^{2}$. With the SHA-256 core, \XT requires 0.0014 $mm^{2}$ area to implement in 7nm process technology, which is only 0.04\% {the chip area} of a contemporary CPU designed at 7nm~\cite{zen2, suggs2020zen2}.

\section{Related Work}
\label{sec:related-work}
To our knowledge, {this is the first work to \One demonstrate that quadruple row activation (\X) in DRAM chips leads to random values by inducing metastability in DRAM sense amplifiers, \Two exploit this phenomenon to design a new true random number generator, \XT{}.}
We have already extensively {compared \XT} to two state{-}of{-}the{-}art high-throughput TRNG designs{~\cite{talukder2019exploiting,kim2019drange}} in {Section}~\ref{sec:comparison-high-throughput}. In this section, we describe other related works.

\subsection{Low-throughput DRAM-based TRNGs}
\label{sec:comparison-lowtput}
\noindent\textbf{Pyo {et al.}~\cite{pyo2009dram}}
{(Table~\ref{table:comparison_rngs}, Pyo+)} generate random numbers using the unpredictability in DRAM command schedule as the entropy source. We calculate the peak theoretical throughput for Pyo+ as 2.17 Mb/s from the number of CPU cycles (45000) that it takes to obtain an 8-bit random number for the system we describe in Section~\ref{sec:ramulator-analysis}. {W}e find the latency of obtaining a 256-bit random number to be 112.5us. 

{\noindent\textbf{Retention-based TRNGs~\cite{sutar2016d,keller2014dynamic}}}
{\One pause DRAM refresh to accumulate a sufficient amount of retention failures~\cite{khan2014efficacy} that is used as the entropy source for true random number generation, \Two read the portion of the DRAM array that contains the retention failures{,} and {\Three} post-process the read {data} using hash functions (e.g., SHA-256) to finally obtain a random number.} 

D-PUF~\cite{sutar2016d} {(Table~\ref{table:comparison_rngs}, D-PUF)} partitions the DRAM into 4 MiB large regions and pauses DRAM refresh for 40 seconds for a region to accumulate a sufficient amount of retention failures in DRAM. {D-PUF uses the SHA-256 hash function to post-process {the} data read from each region to generate a 256-bit random number.} This incurs a minimum latency of 40 seconds to generate random numbers. We optimistically calculate the throughput of {D-PUF} assuming a four-channel system with 128 GiBs of DRAM. We also ignore the time it takes to read out 128 GiBs of data. When 1\% of available DRAM {(i.e., approximately \emph{327} 4 MiB large regions)} is reserved for retention failures, D-PUF's TRNG throughput is 0.002 Mb/s. Even when all {DRAM (32K regions)} is used, D-PUF can achieve {only} 0.20 Mb/s peak throughput. 

{Keller+~\cite{keller2014dynamic} {(Table~\ref{table:comparison_rngs}, Keller+)} partitions the DRAM into 1 MiB large regions and pauses DRAM refresh for 320 seconds. Following an analysis similar to ours on D-PUF~\cite{sutar2016d}, we find Keller+'s TRNG latency for a 256-bit random number to be 320 seconds and its TRNG throughput to be {only} 0.025 Mb/s, assuming a four-channel system with 128 GiB DRAM fully utilized for true random number generation.}

{\noindent\textbf{Startup {v}alue-{b}ased TRNGs~\cite{eckert2017drng}}} {(Table~\ref{table:comparison_rngs}, DRNG)}
{use the startup values in DRAM cells that are accessed immediately after a DRAM device is powered up.}
{These TRNGs} \emph{cannot} be used as a streaming true random number source as {they} require a DRAM power cycle to generate random numbers. We estimate the minimum latency of {this category of TRNGs} from the time it takes to execute a DDR4 power-up initialization sequence~\cite{ddr4operationhynix}, which is \SI{700}{\micro\second}.

{All t}hese DRAM-based TRNGs provide very low random number {generation} throughput and incur high latency. Low-throughput TRNGs are unlikely to be useful in satisfying today's workloads with high throughput random number requirements (e.g., machine learning, cryptography, simulations~\cite{gutterman2006analysis, von2007dual, kim2017nano, drutarovsky2007robust, kwok2006fpga, cherkaoui2013very, zhang2017high, quintessence2015white,clarke2011robust,lu2015fpga,hull1962random, ma2016quantum, botha2005gammaray, davis1956some,zhang2016survey,mich2010machine,schmidt1992feedforward,darrell1998genetic,stipvcevic2014true}). \XT, on the other hand, can satisfy the high-throughput requirements of these workloads.

\vspace{1mm}
\subsection{{Non-DRAM-based} TRNGs That Require\\Specialized Hardware}
\vspace{1mm}
Many prior works design high-throughput TRNGs that are based on specialized hardware~\cite{gehring2020ultra,zhang2017gbit,amaki2015oscillator, yang2016all, bucci2003high,bhargava2015robust,petrie2000noise,mathew20122, brederlow2006low, tokunaga2008true, kinniment2002design, holleman20083, holcomb2009power, wang2016gbps,pareschi2006fast,stefanov2000optical,restituto1993nonlinear}. Unfortunately, it is costly to integrate these substrates into {especially low-cost} commodity systems {as well as future processing-in-memory systems} for true random number generation. 
{Existing TRNGs in some commodity systems~\cite{arm2020true,jun1999intel,amd2017random} both (i) consume die area to implement specialized circuitry {(e.g., ring oscillators~\cite{ning2015design})} that harnesses entropy from physical phenomena and (ii) are limited in throughput. For example, the TRNG in a recent high-end AMD Zen3 processor can provide up to 3.{18} Gb/s throughput {per core, assuming a 4 GHz clock rate}~\cite{fog21instruction}, which is {only} {{23.11}}\% of the throughput QUAC-TRNG can provide {(on a four-channel DDR4-2400 system)}.} 

{In general, choosing a TRNG is a design-time decision that requires balancing needs with costs. QUAC-TRNG provides high-throughput true random number generation without introducing dedicated hardware for TRNGs. Instead, QUAC-TRNG leverages {widely-used} commodity DRAM as an entropy source. Therefore, QUAC-TRNG offers a new design point that can enable new applications that were previously infeasible with alternative TRNGs, especially for systems where the costs of on-chip TRNGs may be prohibitive (e.g., heavily constrained embedded systems, processing-in-memory architectures). For example, QUAC-TRNG would enable {processing-in-memory} systems~\cite{mutlu2020modern, seshadri2017ambit, Hajinazar2021SIMDRAM, upmem2018} to execute security workloads as it enables true random number generation directly within {a DRAM chip}.}

\subsection{Multiple Row Activation In DRAM}
{\textbf{A{mbit}~{\cite{seshadri2017ambit}} and ComputeDRAM~\cite{fei2019computedram}.}}
{{Seshadri et al.~\cite{seshadri2015fast, seshadri2017ambit,seshadri2016buddy,seshadri2020indram} {introduce} the idea of triple row activation in DRAM, showing that this operation leads to a bitwise majority function across the three activated rows. ComputeDRAM~\cite{fei2019computedram} shows that a similar behavior can be observed in real {off-the-shelf} DRAM chips by carefully {reducing} the timing parameters between consecutive {DRAM} commands.}} {We build on these works and introduce quadruple activation (\X{}), which leads to a fundamentally different phenomenon {on real {off-the-shelf} DRAM chips}, i.e., simultaneous activation of four DRAM rows. We exploit this phenomenon to generate true random numbers at high-throughput and low-latency.}

\textbf{CROW~\cite{CROW} {and MCR-DRAM~\cite{choi2015multiple}}} propose {a DRAM-based substrate to simultaneously activate multiple DRAM} rows {with the same data content} to reduce access latency.
{\textbf{RowClone~\cite{seshadri2013rowclone}}}
{enables \emph{consecutive} activation of two DRAM rows to copy data in DRAM. These mechanisms \One require changes to DRAM chips and \Two do not generate random numbers.}

\section{Conclusion}
\label{sec:conclusion}
We {introduce} \XT, a high-throughput and low-latency DRAM-based TRNG that can be implemented in commodity systems at low cost. The key idea of \XT is to induce metastability on many DRAM sense amplifiers in parallel by {exploiting a phenomenon {we observe}, quadruple row activation (QUAC), which} simultaneously activates four DRAM rows {in real DRAM chips}. {Via a detailed characterization of 136 real DRAM chips, w}e show that \XT{} {produces random bitstreams that pass all 15 NIST STS tests, and} generates high-quality true random numbers at {3.44} Gb/s throughput.
We {compare \XT{} against} prior work {that we evaluate} under two configurations, bas{ic} (as proposed) and enhanced (throughput-optimized). \XT outperforms the state{-}of{-}the{-}art DRAM-based TRNG in throughput by {15.08$\times$} and {1.41$\times$} for the bas{ic} and the enhanced configurations, respectively. \XT scales well with DRAM bandwidth and outperforms the enhanced version of the state-of-the-art {by} 2.03$\times$ at {projected} future DRAM transfer rates {(12 GT/s)}. We conclude that {\XT reliably generates true random numbers at high-throughput and low-latency in real DRAM chips.}

\section*{Acknowledgements}
\label{sec:acknowledgements}
\vspace{2mm}
{We thank the anonymous reviewers of ISCA 2021 for feedback and the SAFARI group members for feedback and the stimulating intellectual environment they provide. We acknowledge the generous gifts provided by our industrial partners: Google, Huawei, Intel, Microsoft, and VMware.}
\vspace{2mm}

\bibliographystyle{IEEEtranS}
\setstretch{0.979}
\bibliography{main,drange}

\begin{thebibliography}{100}
\providecommand{\url}[1]{#1}
\csname url@samestyle\endcsname
\providecommand{\newblock}{\relax}
\providecommand{\bibinfo}[2]{#2}
\providecommand{\BIBentrySTDinterwordspacing}{\spaceskip=0pt\relax}
\providecommand{\BIBentryALTinterwordstretchfactor}{4}
\providecommand{\BIBentryALTinterwordspacing}{\spaceskip=\fontdimen2\font plus
\BIBentryALTinterwordstretchfactor\fontdimen3\font minus
  \fontdimen4\font\relax}
\providecommand{\BIBforeignlanguage}[2]{{%
\expandafter\ifx\csname l@#1\endcsname\relax
\typeout{** WARNING: IEEEtranS.bst: No hyphenation pattern has been}%
\typeout{** loaded for the language `#1'. Using the pattern for}%
\typeout{** the default language instead.}%
\else
\language=\csname l@#1\endcsname
\fi
#2}}
\providecommand{\BIBdecl}{\relax}
\BIBdecl

\bibitem{cacti}
``{CACTI: An integrated cache and memory access time, cycle time, area,
  leakage, and dynamic power model},''
  \url{https://www.hpl.hp.com/research/cacti/}.

\bibitem{rambusmodel}
``{DRAM Power Model},'' \url{https://www.rambus.com/energy/}.

\bibitem{helionipcores}
``{Fast Hashing Cores},'' \url{https://www.heliontech.com/fast\_hash.htm}.

\bibitem{ramulatorgithub}
``{Ramulator Source Code},'' \url{https://github.com/CMU-SAFARI/ramulator}.

\bibitem{aga2017compute}
S.~Aga \emph{et~al.}, ``{Compute Caches},'' in \emph{HPCA}, 2017.

\bibitem{ahn2015scalable}
J.~Ahn \emph{et~al.}, ``{A Scalable Processing-in-Memory Accelerator for
  Parallel Graph Processing},'' in \emph{{ISCA}}, 2015.

\bibitem{ahn2015pim}
J.~Ahn \emph{et~al.}, ``{PIM-Enabled Instructions: a Low-overhead,
  Locality-aware Processing-in-Memory Architecture},'' in \emph{{ISCA}}, 2015.

\bibitem{akin2015data}
B.~Akin \emph{et~al.}, ``{Data Reorganization in Memory Using 3D-Stacked
  DRAM},'' in \emph{ISCA}, 2015.

\bibitem{amaki2015oscillator}
T.~Amaki \emph{et~al.}, ``{An Oscillator-based True Random Number Generator
  with Process and Temperature Tolerance},'' in \emph{DAC}, 2015.

\bibitem{amd2017random}
AMD, ``{AMD Random Number Generator},''
  \url{https://www.amd.com/system/files/TechDocs/amd-random-number-generator.pdf}.

\bibitem{zen2}
AMD, ``{AMD Zen2 Microarchitecture},''
  \url{https://en.wikichip.org/wiki/amd/microarchitectures/zen_2}.

\bibitem{arm2020true}
ARM, ``{ARM True Random Number Generator (TRNG) Technical Reference Manual
  Revision r0p0},''
  \url{https://developer.arm.com/documentation/100976/0000/Introduction/Features}.

\bibitem{babarinsa2015jafar}
O.~O. Babarinsa and S.~Idreos, ``{JAFAR: Near-Data Processing for Databases},''
  in \emph{{SIGMOD}}, 2015.

\bibitem{bagini1999design}
V.~Bagini and M.~Bucci, ``{A Design of Reliable True Random Number Generator
  for Cryptographic Applications},'' in \emph{CHES}, 1999.

\bibitem{talukder2019exploiting}
B.~M.~S. {Bahar Talukder} \emph{et~al.}, ``{Exploiting DRAM Latency Variations
  for Generating True Random Numbers},'' in \emph{ICCE}, 2019.

\bibitem{bakiri2018survey}
M.~Bakiri \emph{et~al.}, ``{Survey on Hardware Implementation of Random Number
  Generators on FPGA: Theory and Experimental Analyses},'' \emph{{CSR}}, 2018.

\bibitem{baldanzi2020cryptographically}
L.~Baldanzi \emph{et~al.}, ``{Cryptographically Secure Pseudo-Random Number
  Generator IP-Core Based on SHA2 Algorithm},'' \emph{Sensors}, 2020.

\bibitem{barangi2016straintronics}
M.~Barangi \emph{et~al.}, ``{Straintronics-Based True Random Number Generator
  for High-Speed and Energy-Limited Applications},'' in \emph{IEEE Trans.
  Magn}, 2016.

\bibitem{barenghi2018software}
A.~Barenghi \emph{et~al.}, ``{Software-Only Reverse Engineering of Physical
  DRAM Mappings for Rowhammer Attacks},'' in \emph{IVSW}, 2018.

\bibitem{rukhin2001statistical}
L.~Bassham \emph{et~al.}, ``{A Statistical Test Suite for Random and
  Pseudorandom Number Generators for Cryptographic Applications},'' Special
  Publication (NIST SP), 2010.

\bibitem{bhargava2015robust}
M.~Bhargava \emph{et~al.}, ``{Robust True Random Number Generator Using
  Hot-Carrier Injection Balanced Metastable Sense Amplifiers},'' in
  \emph{HOST}, 2015.

\bibitem{boroumand2021mitigating}
A.~Boroumand \emph{et~al.}, ``{Mitigating Edge Machine Learning Inference
  Bottlenecks: An Empirical Study on Accelerating Google Edge Models},''
  {arXiv:2103.00768}, 2021.

\bibitem{boroumand2018google}
A.~Boroumand \emph{et~al.}, ``{Google Workloads for Consumer Devices:
  Mitigating Data Movement Bottlenecks},'' in \emph{ASPLOS}, 2018.

\bibitem{boroumand2017lazypim}
A.~Boroumand \emph{et~al.}, ``{LazyPIM: An Efficient Cache Coherence Mechanism
  for Processing-in-Memory},'' in \emph{{CAL}}, 2017.

\bibitem{boroumand2021polynesia}
A.~Boroumand \emph{et~al.}, ``Polynesia: Enabling effective hybrid
  transactional/analytical databases with specialized hardware/software
  co-design,'' {arXiv:2103.00798}, 2021.

\bibitem{boroumand2019conda}
A.~Boroumand \emph{et~al.}, ``{CONDA: Efficient Cache Coherence Support for
  Near-Data Accelerators},'' in \emph{ISCA}, 2019.

\bibitem{botha2005gammaray}
R.~Botha, ``{The Development of a Hardware Random Number Generator for
  Gamma-ray Astronomy},'' {PhD Dissertation}, North-West University, 2005.

\bibitem{brederlow2006low}
R.~Brederlow \emph{et~al.}, ``{A Low-power True Random Number Generator using
  Random Telegraph Noise of Single Oxide-traps},'' in \emph{ISSCC}, 2006.

\bibitem{bucci2003high}
M.~Bucci \emph{et~al.}, ``{A High-speed Oscillator-based Truly Random Number
  Source for Cryptographic Applications on a Smart Card IC},'' in \emph{TC},
  2003.

\bibitem{chandrasekar2014exploiting}
K.~Chandrasekar \emph{et~al.}, ``{Exploiting Expendable Process-Margins in
  DRAMs for Run-Time Performance Optimization},'' in \emph{DATE}, 2014.

\bibitem{chang2017thesis}
K.~K. Chang, ``{Understanding and Improving Latency of DRAM-Based Memory
  Systems},'' {PhD Dissertation}, Carnegie Mellon University, 2017.

\bibitem{chang2016understanding}
K.~K. Chang \emph{et~al.}, ``{Understanding Latency Variation in Modern DRAM
  Chips: Experimental Characterization, Analysis, and Optimization},'' in
  \emph{SIGMETRICS}, 2016.

\bibitem{chang2014improving}
K.~K. Chang \emph{et~al.}, ``{Improving DRAM Performance by Parallelizing
  Refreshes with Accesses},'' in \emph{HPCA}, 2014.

\bibitem{chang2016low}
K.~K. Chang \emph{et~al.}, ``{Low-Cost Inter-Linked Subarrays (LISA): Enabling
  Fast Inter-Subarray Data Movement in DRAM},'' in \emph{{HPCA}}, 2016.

\bibitem{chang_understanding2017}
K.~K. Chang \emph{et~al.}, ``{Understanding Reduced-Voltage Operation in Modern
  DRAM Devices: Experimental Characterization, Analysis, and Mechanisms},'' in
  \emph{SIGMETRICS}, 2017.

\bibitem{chatterjee2017architecting}
N.~Chatterjee \emph{et~al.}, ``{Architecting an Energy-Efficient DRAM System
  for GPUs},'' in \emph{HPCA}, 2017.

\bibitem{cherkaoui2013very}
A.~Cherkaoui \emph{et~al.}, ``{A Very High Speed True Random Number Generator
  with Entropy Assessment},'' in \emph{CHES}, 2013.

\bibitem{chi2016prime}
P.~Chi \emph{et~al.}, ``{PRIME: A Novel Processing-in-Memory Architecture for
  Neural Network Computation in ReRAM-Based Main Memory},'' in \emph{ISCA},
  2016.

\bibitem{choi2015multiple}
J.~Choi \emph{et~al.}, ``{Multiple Clone Row DRAM: A Low Latency and Area
  Optimized DRAM},'' in \emph{ISCA}, 2015.

\bibitem{clarke2011robust}
P.~J. Clarke \emph{et~al.}, ``{Robust Gigahertz Fiber Quantum Key
  Distribution},'' \emph{Applied Physics Letters}, 2011.

\bibitem{cojocar2020rowhammer}
L.~Cojocar \emph{et~al.}, ``{Are We Susceptible to Rowhammer? An End-to-End
  Methodology for Cloud Providers},'' in \emph{S\&P}, 2020.

\bibitem{davis1956some}
P.~Davis and P.~Rabinowitz, ``{Some Monte Carlo Experiments in Computing
  Multiple Integrals},'' \emph{Mathematical Tables and Other Aids to
  Computation}, 1956.

\bibitem{restituto1993nonlinear}
M.~{Degaldo-Restituto} \emph{et~al.}, ``{Nonlinear switched-current CMOS IC for
  random signal generation},'' \emph{Electronics Letters}, 1993.

\bibitem{devaux2019true}
F.~Devaux, ``{The True Processing in Memory Accelerator},'' in \emph{HCS},
  2019.

\bibitem{lee2016phdthesis}
{Donghyuk Lee}, ``{Reducing DRAM Latency at Low Cost by Exploiting
  Heterogeneity},'' {PhD Dissertation}, {Carnegie Mellon University}, 2016.

\bibitem{drutarovsky2007robust}
M.~Drutarovsky and P.~Galajda, ``{A Robust Chaos-based True Random Number
  Generator Embedded in Reconfigurable Switched-Capacitor Hardware},'' in
  \emph{Radioelektronika}, 2007.

\bibitem{eckert2017drng}
C.~Eckert \emph{et~al.}, ``{DRNG: DRAM-based Random Number Generation Using its
  Startup Value Behavior},'' in \emph{MWSCAS}, 2017.

\bibitem{farmahini2015nda}
A.~Farmahini-Farahani \emph{et~al.}, ``{NDA: Near-DRAM Acceleration
  Architecture Leveraging Commodity DRAM Devices and Standard Memory
  Modules},'' in \emph{{HPCA}}, 2015.

\bibitem{fernandez2020natsa}
I.~Fernandez \emph{et~al.}, ``{NATSA: A Near-Data Processing Accelerator for
  Time Series Analysis},'' 2020.

\bibitem{fips2012180}
{FIPS, PUB}, ``{180-2: Secure hash standard (SHS)},'' \emph{US Department of
  Commerce, National Institute of Standards and Technology (NIST)}, 2012.

\bibitem{fog21instruction}
A.~Fog, ``{Lists of instruction latencies, throughputs and micro-operation
  breakdowns for Intel, AMD, and VIA CPUs},''
  \url{https://www.agner.org/optimize/instruction_tables.pdf}.

\bibitem{frigo2020trrespass}
P.~Frigo \emph{et~al.}, ``{TRRespass: Exploiting the Many Sides of Target Row
  Refresh},'' in \emph{{S\&P}}, 2020.

\bibitem{fei2019computedram}
F.~Gao \emph{et~al.}, ``{ComputeDRAM: In-Memory Compute Using Off-the-Shelf
  DRAMs},'' in \emph{MICRO}, 2019.

\bibitem{gao2015practical}
M.~Gao \emph{et~al.}, ``{Practical Near-Data Processing for In-Memory Analytics
  Frameworks},'' in \emph{{PACT}}, 2015.

\bibitem{gao2016hrl}
M.~Gao and C.~Kozyrakis, ``{HRL: Efficient and Flexible Reconfigurable Logic
  for Near-Data Processing},'' in \emph{{HPCA}}, 2016.

\bibitem{gehring2020ultra}
T.~Gehring \emph{et~al.}, ``{Ultra-Fast Real-Time Quantum Random Number
  Generator with Correlated Measurement Outcomes and Rigorous Security
  Certification},'' {arXiv:1812.05377}, 2020.

\bibitem{ghose2019processing}
S.~Ghose \emph{et~al.}, ``{Processing-in-Memory: A Workload-Driven
  Perspective},'' \emph{IBM JRD}, 2019.

\bibitem{ghose2018enabling}
S.~Ghose \emph{et~al.}, ``{Enabling the Adoption of Processing-in-Memory:
  Challenges, Mechanisms, Future Research Directions},''
  \emph{arXiv:1802.00320}, 2018.

\bibitem{giannoula2021syncron}
C.~Giannoula \emph{et~al.}, ``{SynCron: Efficient Synchronization Support for
  Near-Data-Processing Architectures},'' in \emph{HPCA}, 2021.

\bibitem{juan2021benchmarking}
J.~G{\'o}mez-Luna \emph{et~al.}, ``{Benchmarking a New Paradigm: Understanding
  a Modern Processing-in-Memory Architecture},'' {arXiv:2105.03814}, 2021.

\bibitem{gutterman2006analysis}
Z.~Gutterman \emph{et~al.}, ``{Analysis of the Linux Random Number
  Generator},'' in \emph{SP}, 2006.

\bibitem{Hajinazar2021SIMDRAM}
N.~Hajinazar \emph{et~al.}, ``{SIMDRAM}: {A Framework for Bit-Serial SIMD
  Processing Using DRAM},'' in \emph{ASPLOS}, 2021.

\bibitem{hashemian2015robust}
M.~S. Hashemian \emph{et~al.}, ``{A Robust Authentication Methodology Using
  Physically Unclonable Functions in DRAM Arrays},'' in \emph{DATE}, 2015.

\bibitem{SoftMC}
H.~{Hassan} \emph{et~al.}, ``{SoftMC: A Flexible and Practical Open-Source
  Infrastructure for Enabling Experimental DRAM Studies},'' in \emph{HPCA},
  2017.

\bibitem{CROW}
H.~Hassan \emph{et~al.}, ``{CROW: A Low-Cost Substrate for Improving DRAM
  Performance, Energy Efficiency, and Reliability},'' in \emph{ISCA}, 2019.

\bibitem{hassan2015near}
S.~M. Hassan \emph{et~al.}, ``{Near Data Processing: Impact and Optimization of
  3D Memory System Architecture on the Uncore},'' in \emph{{MEMSYS}}, 2015.

\bibitem{holcomb2007initial}
D.~E. Holcomb \emph{et~al.}, ``{Initial SRAM State as a Fingerprint and Source
  of True Random Numbers for RFID Tags},'' in \emph{RFID}, 2007.

\bibitem{holcomb2009power}
D.~E. Holcomb \emph{et~al.}, ``{Power-Up SRAM State as an Identifying
  Fingerprint and Source of True Random Numbers},'' \emph{ToC}, 2009.

\bibitem{holleman20083}
J.~Holleman \emph{et~al.}, ``{A 3mu W CMOS True Random Number Generator with
  Adaptive Floating-Gate Offset Cancellation},'' \emph{JSSC}, 2008.

\bibitem{horiguchi1997redundancy}
M.~Horiguchi, ``{Redundancy Techniques for High-Density DRAMs},'' in
  \emph{ISIS}, 1997.

\bibitem{hsieh2016transparent}
K.~Hsieh \emph{et~al.}, ``{Transparent Offloading and Mapping (TOM): Enabling
  Programmer-Transparent Near-Data Processing in GPU Systems},'' in
  \emph{{ISCA}}, 2016.

\bibitem{hsieh2016accelerating}
K.~Hsieh \emph{et~al.}, ``{Accelerating Pointer Chasing in 3D-Stacked Memory:
  Challenges, Mechanisms, Evaluation},'' in \emph{{ICCD}}, 2016.

\bibitem{hull1962random}
T.~E. Hull and A.~R. Dobell, ``{Random Number Generators},'' \emph{SIAM
  Review}, 1962.

\bibitem{humood2021DTRNG}
K.~Humood \emph{et~al.}, ``{{DTRNG}}: {{Low Cost}} and {{Robust True Random
  Number Generator Using DRAM Weak Write Scheme}},'' in \emph{{ISCAS}}, 2021.

\bibitem{itoh2013vlsi}
K.~Itoh, \emph{{VLSI Memory Chip Design}}.\hskip 1em plus 0.5em minus
  0.4em\relax Springer, 2001.

\bibitem{jedecDDR4}
JEDEC, ``{DDR4},'' \emph{JEDEC Standard JESD79--4}, 2012.

\bibitem{gddr5}
{JEDEC}, ``{Graphics Double Data Rate (GDDR5) SGRAM Standard},'' 2016.

\bibitem{jun1999intel}
B.~Jun and P.~Kocher, ``{The Intel Random Number Generator (White Paper)},''
  \emph{Cryptography Research Inc.}, 1999.

\bibitem{kang2014co}
U.~Kang \emph{et~al.}, ``{Co-Architecting Controllers and DRAM to Enhance DRAM
  Process Scaling},'' in \emph{The Memory Forum}, 2014.

\bibitem{keeth2001dram}
B.~Keeth and R.~Baker, \emph{{DRAM Circuit Design: A Tutorial}}.\hskip 1em plus
  0.5em minus 0.4em\relax Wiley, 2001.

\bibitem{keller2014dynamic}
C.~Keller \emph{et~al.}, ``{Dynamic Memory-based Physically Unclonable Function
  for the Generation of Unique Identifiers and True Random Numbers},'' in
  \emph{ISCAS}, 2014.

\bibitem{khan2014efficacy}
S.~Khan \emph{et~al.}, ``{The Efficacy of Error Mitigation Techniques for DRAM
  Retention Failures: A Comparative Experimental Study},'' in
  \emph{SIGMETRICS}, 2014.

\bibitem{khan2016parbor}
S.~Khan \emph{et~al.}, ``{PARBOR: An Efficient System-Level Technique to Detect
  Data-Dependent Failures in DRAM},'' in \emph{DSN}, 2016.

\bibitem{khan2017detecting}
S.~Khan \emph{et~al.}, ``{Detecting and Mitigating Data-Dependent DRAM Failures
  by Exploiting Current Memory Content},'' in \emph{MICRO}, 2017.

\bibitem{kim2017nano}
J.~{Kim} \emph{et~al.}, ``{Nano-Intrinsic True Random Number Generation: A
  Device to Data Study},'' \emph{IEEE TCAS}, 2019.

\bibitem{kim2020phdthesis}
J.~S. Kim, ``{Improving DRAM Performance, Security, and Reliability by
  Understanding and Exploiting DRAM Timing Parameter Margins},'' {PhD
  Dissertation}, {Carnegie Mellon University}, 2020.

\bibitem{kim2018latpuf}
J.~S. {Kim} \emph{et~al.}, ``{The DRAM Latency PUF: Quickly Evaluating Physical
  Unclonable Functions by Exploiting the Latency-Reliability Tradeoff in Modern
  Commodity DRAM Devices},'' in \emph{HPCA}, 2018.

\bibitem{kim2019drange}
J.~S. {Kim} \emph{et~al.}, ``{D-RaNGe: Using Commodity DRAM Devices to Generate
  True Random Numbers with Low Latency and High Throughput},'' in \emph{HPCA},
  2019.

\bibitem{kim2018grim}
J.~S. Kim \emph{et~al.}, ``{GRIM-Filter: Fast Seed Location Filtering in DNA
  Read Mapping Using Processing-in-memory Technologies},'' \emph{BMC Genomics},
  2018.

\bibitem{kim2018solar}
J.~S. Kim \emph{et~al.}, ``{Solar-DRAM: Reducing DRAM Access Latency by
  Exploiting the Variation in Local Bitlines},'' in \emph{ICCD}, 2018.

\bibitem{kim2020revisiting}
J.~S. Kim \emph{et~al.}, ``{Revisiting RowHammer: An Experimental Analysis of
  Modern DRAM Devices and Mitigation Techniques},'' in \emph{ISCA}, 2020.

\bibitem{kim2014flipping}
Y.~Kim \emph{et~al.}, ``{Flipping Bits in Memory Without Accessing Them: An
  Experimental Study of DRAM Disturbance Errors},'' in \emph{ISCA}, 2014.

\bibitem{kim2012case}
Y.~Kim \emph{et~al.}, ``{A Case for Exploiting Subarray-level Parallelism
  (SALP) in DRAM},'' in \emph{ISCA}, 2012.

\bibitem{kim2016ramulator}
Y.~Kim \emph{et~al.}, ``{Ramulator: A Fast and Extensible DRAM Simulator},'' in
  \emph{CAL}, 2016.

\bibitem{kinniment2002design}
D.~Kinniment and E.~Chester, ``{Design of an On-chip Random Number Generator
  using Metastability},'' in \emph{ESSCIRC}, 2002.

\bibitem{kocc2009cryptographic}
{\c{C}}.~K. Ko{\c{c}}, ``{About Cryptographic Engineering},'' in
  \emph{Cryptographic Engineering}, 2009.

\bibitem{kwok2006fpga}
S.~H. Kwok and E.~Y. Lam, ``{FPGA-based High-speed True Random Number Generator
  for Cryptographic Applications},'' in \emph{TENCON}, 2006.

\bibitem{kwon202125}
Y.-C. Kwon \emph{et~al.}, ``{25.4 A 20nm 6GB Function-In-Memory DRAM, Based on
  HBM2 with a 1.2 TFLOPS Programmable Computing Unit Using Bank-Level
  Parallelism, for Machine Learning Applications},'' in \emph{ISSCC}, 2021.

\bibitem{lee2016reducing}
D.~Lee, ``{Reducing DRAM Latency at Low Cost by Exploiting Heterogeneity},''
  {PhD Dissertation}, Carnegie Mellon University, 2016.

\bibitem{lee2015simultaneous}
D.~Lee \emph{et~al.}, ``{Simultaneous Multi-Layer Access: Improving 3D-Stacked
  Memory Bandwidth at Low Cost},'' in \emph{{TACO}}, 2016.

\bibitem{lee2017design}
D.~Lee \emph{et~al.}, ``{{Design-Induced Latency Variation in Modern DRAM
  Chips: Characterization, Analysis, and Latency Reduction Mechanisms}},'' in
  \emph{SIGMETRICS}, 2017.

\bibitem{aldram}
D.~Lee \emph{et~al.}, ``{Adaptive-Latency DRAM: Optimizing DRAM Timing for the
  Common-Case},'' in \emph{HPCA}, 2015.

\bibitem{li2016pinatubo}
S.~Li \emph{et~al.}, ``{Pinatubo: A Processing-in-Memory Architecture for Bulk
  Bitwise Operations in Emerging Non-Volatile Memories},'' in \emph{DAC}, 2016.

\bibitem{zhang2017gbit}
Z.~Limeng \emph{et~al.}, ``{640-Gbit/s Fast Physical Random Number Generation
  Using a Broadband Chaotic Semiconductor Laser},'' \emph{Scientific Reports},
  2017.

\bibitem{liu2013experimental}
J.~Liu \emph{et~al.}, ``{An Experimental Study of Data Retention Behavior in
  Modern DRAM Devices: Implications for Retention Time Profiling Mechanisms},''
  in \emph{ISCA}, 2013.

\bibitem{raidr}
J.~Liu \emph{et~al.}, ``{RAIDR: Retention-Aware Intelligent DRAM Refresh},'' in
  \emph{ISCA}, 2012.

\bibitem{liu2017concurrent}
Z.~Liu \emph{et~al.}, ``{Concurrent Data Structures for Near-Memory
  Computing},'' in \emph{{SPAA}}, 2017.

\bibitem{lu2015fpga}
X.~Lu \emph{et~al.}, ``{FPGA Based Digital Phase-coding Quantum Key
  Distribution System},'' \emph{Science China Physics, Mechanics \& Astronomy},
  2015.

\bibitem{ma2016quantum}
X.~Ma \emph{et~al.}, ``{Quantum Random Number Generation},'' \emph{Quantum
  Inf.}, 2016.

\bibitem{mathew20122}
S.~K. Mathew \emph{et~al.}, ``{2.4 Gbps, 7 mW All-digital PVT-variation
  Tolerant True Random Number Generator for 45 nm CMOS High-performance
  Microprocessors},'' in \emph{JSSC}, 2012.

\bibitem{mehrotra2001modeling}
V.~Mehrotra, ``{Modeling the Effects of Systematic Process Variation of Circuit
  Performance},'' {PhD Dissertation}, {Massachusetts Institute of Technology},
  2001.

\bibitem{mich2010machine}
Y.~Mich{\'e} \emph{et~al.}, ``{Machine Learning Techniques based on Random
  Projections},'' in \emph{ESANN}, 2010.

\bibitem{morad2015gp}
A.~Morad \emph{et~al.}, ``{GP-SIMD Processing-in-Memory},'' in \emph{{TACO}},
  2015.

\bibitem{mutlu2013memory}
O.~Mutlu, ``{Memory Scaling: A Systems Architecture Perspective},'' in
  \emph{IMW}, 2013.

\bibitem{mutlu2019processing}
O.~Mutlu \emph{et~al.}, ``{Processing Data Where it Makes Sense: Enabling
  In-Memory Computation},'' \emph{Microprocessors and Microsystems}, 2019.

\bibitem{mutlu2020modern}
O.~Mutlu \emph{et~al.}, ``{A Modern Primer on Processing in Memory},''
  {arXiv:2012.03112}, 2020.

\bibitem{ning2015design}
L.~Ning \emph{et~al.}, ``{Design and Validation of High Speed True Random
  Number Generators Based on Prime-length Ring Oscillators},'' \emph{The
  Journal of China Universities of Posts and Telecommunications}, 2015.

\bibitem{oliveira2021new}
G.~F. Oliveira \emph{et~al.}, ``{DAMOV: A New Methodology and Benchmark Suite
  for Evaluating Data Movement Bottlenecks},'' {arXiv:2105.03725}, 2021.

\bibitem{orosa2019dataplant}
L.~Orosa \emph{et~al.}, ``Dataplant: Enhancing system security with low-cost
  in-dram value generation primitives,'' {arXiv:1902.07344}, 2019.

\bibitem{orosa2021codic}
L.~Orosa \emph{et~al.}, ``{CODIC: A Low-cost Substrate for Enabling Custom
  In-DRAM Functionalities and Optimizations},'' in \emph{ISCA}, 2021.

\bibitem{pareschi2006fast}
F.~Pareschi \emph{et~al.}, ``{A Fast Chaos-based True Random Number Generator
  for Cryptographic Applications},'' in \emph{ESSCIRC}, 2006.

\bibitem{patel2020beer}
M.~Patel \emph{et~al.}, ``{Bit-Exact ECC Recovery (BEER): Determining DRAM
  On-Die ECC Functions by Exploiting DRAM Data Retention Characteristics},'' in
  \emph{{MICRO}}, 2020.

\bibitem{patel2017reaper}
M.~Patel \emph{et~al.}, ``{The Reach Profiler (REAPER): Enabling the Mitigation
  of DRAM Retention Failures via Profiling at Aggressive Conditions},'' in
  \emph{ISCA}, 2017.

\bibitem{pattnaik2016scheduling}
A.~Pattnaik \emph{et~al.}, ``{Scheduling Techniques for GPU Architectures with
  Processing-in-Memory Capabilities},'' in \emph{{PACT}}, 2016.

\bibitem{petrie2000noise}
C.~S. Petrie and J.~A. Connelly, ``{A Noise-based IC Random Number Generator
  for Applications in Cryptography},'' in \emph{Trans. Circuits Syst. I}, 2000.

\bibitem{pyo2009dram}
C.~Pyo \emph{et~al.}, ``{DRAM as Source of Randomness},'' in \emph{IET}, 2009.

\bibitem{quintessence2015white}
{Quintessence Labs}, ``{Random Number Generators White Paper},'' 2015.

\bibitem{qureshi2015avatar}
M.~K. Qureshi \emph{et~al.}, ``{AVATAR: A Variable-Retention-Time (VRT) Aware
  Refresh for DRAM Systems},'' in \emph{DSN}, 2015.

\bibitem{rivest1992md5}
R.~Rivest, ``{The MD5 Message-Digest Algorithm},'' in \emph{RFC}, 1992.

\bibitem{rock2005pseudorandom}
A.~R{\"o}ck, ``{Pseudorandom Number Generators for Cryptographic
  Applications},'' Master's thesis, Paris-Lodron-Universit{\"a}t Salzburg,
  2005.

\bibitem{satoh2005ASIC}
A.~{Satoh} and T.~{Inoue}, ``{ASIC Hardware Focused Comparison for Hash
  Functions MD5, RIPEMD-160, and SHS},'' in \emph{{ITCC}}, 2005.

\bibitem{schmidt1992feedforward}
W.~F. {Schmidt} \emph{et~al.}, ``{Feedforward Neural Networks with Random
  Weights},'' in \emph{ICPR}, 1992.

\bibitem{seshadri2016simple}
V.~Seshadri, ``{Simple DRAM and Virtual Memory Abstractions to Enable Highly
  Efficient Memory Systems},'' {PhD Dissertation}, Carnegie Mellon University,
  2016.

\bibitem{seshadri2015fast}
V.~Seshadri \emph{et~al.}, ``{Fast Bulk Bitwise AND and OR in DRAM},''
  \emph{IEEE CAL}, 2015.

\bibitem{seshadri2013rowclone}
V.~Seshadri \emph{et~al.}, ``{RowClone: Fast and Energy-Efficient In-DRAM Bulk
  Data Copy and Initialization},'' in \emph{{MICRO}}, 2013.

\bibitem{seshadri2016buddy}
V.~Seshadri \emph{et~al.}, ``{Buddy-RAM: Improving the Performance and
  Efficiency of Bulk Bitwise Operations Using DRAM},'' {arXiv:1611.09988},
  2016.

\bibitem{seshadri2017ambit}
V.~Seshadri \emph{et~al.}, ``{Ambit: In-Memory Accelerator for Bulk Bitwise
  Operations Using Commodity DRAM Technology},'' in \emph{{MICRO}}, 2017.

\bibitem{seshadri2015gather}
V.~Seshadri \emph{et~al.}, ``{Gather-Scatter DRAM: In-DRAM Address Translation
  to Improve the Spatial Locality of Non-Unit Strided Accesses},'' in
  \emph{{MICRO}}, 2015.

\bibitem{seshadri2017simple}
V.~Seshadri and O.~Mutlu, ``{Simple Operations in Memory to Reduce Data
  Movement},'' in \emph{{Advances in Computers}}, 2017.

\bibitem{seshadri2020indram}
V.~Seshadri and O.~Mutlu, ``{In-DRAM Bulk Bitwise Execution Engine},''
  {arXiv:1905.09822}, 2020.

\bibitem{shannon1948mathematical}
C.~E. Shannon, ``{A Mathematical Theory of Communication},'' \emph{Bell System
  Technical Journal}, 1948.

\bibitem{singh2020near}
G.~{Singh} \emph{et~al.}, ``{NERO: A Near High-Bandwidth Memory Stencil
  Accelerator for Weather Prediction Modeling},'' in \emph{FPL}, 2020.

\bibitem{ddr4operationhynix}
{SK Hynix}, ``{DDR4 SDRAM Device Operation}.''

\bibitem{smith1981laser}
R.~T. Smith \emph{et~al.}, ``{Laser Programmable Redundancy and Yield
  Improvement in a 64K DRAM},'' \emph{JSSC}, 1981.

\bibitem{stefanov2000optical}
A.~Stefanov \emph{et~al.}, ``{Optical Quantum Random Number Generator},'' in
  \emph{J. Mod. Opt}, 2000.

\bibitem{stipvcevic2014true}
M.~Stip{\v{c}}evi{\'c} and {\c{C}}.~K. Ko{\c{c}}, ``{True Random Number
  Generators},'' in \emph{Open Problems in Mathematics and Computational
  Science}, 2014.

\bibitem{suggs2020zen2}
D.~Suggs \emph{et~al.}, ``{The AMD ``Zen 2'' Processor},'' \emph{{Hot Chips}},
  2020.

\bibitem{sura2015data}
Z.~Sura \emph{et~al.}, ``{Data Access Optimization in a Processing-in-Memory
  System},'' in \emph{{CF}}, 2015.

\bibitem{sutar2018d}
S.~Sutar \emph{et~al.}, ``{D-PUF: An Intrinsically Reconfigurable DRAM PUF for
  Device Authentication and Random Number Generation},'' in \emph{TECS}, 2018.

\bibitem{sutar2016d}
S.~Sutar \emph{et~al.}, ``{D-PUF: An Intrinsically Reconfigurable DRAM PUF for
  Device Authentication in Embedded Systems},'' in \emph{CASES}, 2016.

\bibitem{tao2017tvl}
S.~Tao and E.~Dubrova, ``{TVL-TRNG: Sub-Microwatt True Random Number Generator
  Exploiting Metastability in Ternary Valued Latches},'' in \emph{ISMVL}, 2017.

\bibitem{tatar2018defeating}
A.~Tatar \emph{et~al.}, ``{Defeating Software Mitigations Against Rowhammer: A
  Surgical Precision Hammer},'' in \emph{RAID}, 2018.

\bibitem{tehranipoor2016robust}
F.~Tehranipoor \emph{et~al.}, ``{Robust Hardware True Random Number Generators
  using DRAM Remanence Effects},'' in \emph{HOST}, 2016.

\bibitem{tokunaga2008true}
C.~Tokunaga \emph{et~al.}, ``{True Random Number Generator with a
  Metastability-based Quality Control},'' in \emph{JSSC}, 2008.

\bibitem{udipi2010rethinking}
A.~N. Udipi \emph{et~al.}, ``{Rethinking DRAM Design and Organization for
  Energy-Constrained Multi-Cores},'' in \emph{ISCA}, 2010.

\bibitem{ugajin2017real}
K.~Ugajin \emph{et~al.}, ``{Real-time fast physical random number generator
  with a photonic integrated circuit},'' \emph{Optics Express}, 2017.

\bibitem{upmem2018}
UPMEM, ``{Introduction to {UPMEM PIM}. Processing-in-memory {(PIM)} on {DRAM}
  accelerator (White Paper)},'' 2018.

\bibitem{van2012efficient}
V.~van~der Leest \emph{et~al.}, ``{Efficient Implementation of True Random
  Number Generator Based on SRAM PUFs},'' in \emph{Cryptography and Security:
  From Theory to Applications}, 2012.

\bibitem{venkatesan2006retention}
R.~K. Venkatesan \emph{et~al.}, ``{Retention-aware Placement in DRAM (RAPID):
  Software Methods for Quasi-non-volatile DRAM},'' in \emph{HPCA}, 2006.

\bibitem{vincent2010stacked}
P.~Vincent \emph{et~al.}, ``{Stacked Denoising Autoencoders: Learning Useful
  Representations in a Deep Network with a Local Denoising Criterion},''
  \emph{JMLR}, 2010.

\bibitem{von2007dual}
V.~von Kaenel and T.~Takayanagi, ``{Dual True Random Number Generators for
  Cryptographic Applications Embedded on a 200 Million Device Dual CPU SOC},''
  in \emph{CICC}, 2007.

\bibitem{neumann1951various}
J.~von Neumann, ``{Various Techniques Used in Connection with Random Digits},''
  in \emph{Monte Carlo Method}, ser. NBS Applied Mathematics Series, 1951.

\bibitem{wang2016gbps}
X.~{Wang} \emph{et~al.}, ``{10-Gbps True Random Number Generator Accomplished
  in ASIC},'' in \emph{RT}, 2016.

\bibitem{wang2020figaro}
Y.~Wang \emph{et~al.}, ``{FIGARO: Improving System Performance via Fine-Grained
  In-DRAM Data Relocation and Caching},'' in \emph{MICRO}, 2020.

\bibitem{wang2016theory}
Y.~Wang \emph{et~al.}, ``{Theory and Implementation of a Very High Throughput
  True Random Number Generator in Field Programmable Gate Array},''
  \emph{{RSI}}, 2016.

\bibitem{darrell1998genetic}
D.~Whitley, ``{A Genetic Algorithm Tutorial},'' \emph{Statistics and
  Computing}, 1998.

\bibitem{yang2016all}
K.~Yang \emph{et~al.}, ``{An All-digital Edge Racing True Random Number
  Generator Robust Against PVT Variations},'' in \emph{JSSC}, 2016.

\bibitem{zhang2014top}
D.~Zhang \emph{et~al.}, ``{TOP-PIM: Throughput-Oriented Programmable Processing
  in Memory},'' in \emph{{HPDC}}, 2014.

\bibitem{zhang2016survey}
L.~Zhang and P.~Suganthan, ``{A Survey of Randomized Algorithms for Training
  Neural Networks},'' \emph{Information Sciences}, 2016.

\bibitem{zhang2017high}
T.~Zhang \emph{et~al.}, ``{High-speed True Random Number Generation Based on
  Paired Memristors for Security Electronics},'' \emph{Nanotechnology}, 2017.

\end{thebibliography}

\doublespacing
\clearpage
\onecolumn
\chapter{\textbf{A. Appendix}}

\begin{table*}[h]
  \centering
  \caption{Sample population of 17 DDR4 modules}
  \resizebox{0.99 \linewidth}{!}{
    \begin{tabular}{rccccccccc}
        \toprule
        \mr{3.5}{\emph{Module}} & \mr{3.5}{\emph{Module Identifier}} & \mr{3.5}{\emph{Chip Identifier}} &
        \mr{3.5}{\emph{\makecell{Freq. \\(MT/s)}}} & \multicolumn{3}{c}{\emph{Organization}} & \multicolumn{3}{c}{\emph{Segment Entropy}}\\
        \cmidrule(lr){5-7}\cmidrule(lr){8-10}
        \multicolumn{4}{c}{} &
        \emph{\makecell{Size\\(GB)}} & \emph{Chips} & \emph{Pins} & \emph{Avg.} & \emph{Max.$^{\dagger{}}$} & \emph{\makecell{Avg.\\(after 30 days)}}\\
        \midrule

        \stripe
        $M1$    & \emph{Unknown} & H5AN4G8NAFR-TFC & 2133  & 4     & 8     & x8 & 1688.1 & 2247.4 & -- \\
        $M2$    & \emph{Unknown} & \emph{Unknown} & 2133  & 4     & 8     & x8 & 1180.4 & 1406.1 & -- \\ 
        \stripe
        $M3$   & \emph{Unknown} & H5AN4G8NAFR-TFC & 2133  & 4     & 8     & x8 & 1205.0 & 1858.3 & 1192.9 \\
        $M4$   & 76TT21NUS1R8-4G & H5AN4G8NAFR-TFC & 2133  & 4     & 8     & x8 & 1608.1 & 2406.5 & 1588.0 \\
        \stripe
        $M5$  & \emph{Unknown} & T4D5128HT-21 & 2133  & 4     & 8     & x8 & 1618.2 & 2121.6 & -- \\

        $M6$   & TLRD44G2666HC18F-SBK & H5AN4G8NMFR-VKC & 2666  &  4     & 8     & x8 & 1211.5 &1444.6 & -- \\
        \stripe
        $M7$  & TLRD44G2666HC18F-SBK &  H5AN4G8NMFR-VKC & 2666  & 4     & 8     & x8 & 1177.7 &1404.4 & --\\
        $M8$ & TLRD44G2666HC18F-SBK &  H5AN4G8NMFR-VKC & 2666  & 4     & 8     & x8  &1332.9 & 1600.9 & 1407.0 \\
        \stripe
        $M9$ & TLRD44G2666HC18F-SBK &  H5AN4G8NMFR-VKC &  2666  & 4     & 8     & x8 & 1137.1 & 1370.9 & --\\

        $M10$  & TLRD44G2666HC18F-SBK &  H5AN4G8NMFR-VKC  &  2666  & 4     & 8     & x8 & 1208.5 & 1473.2 & 1251.8 \\
        \stripe
        $M11$  & TLRD44G2666HC18F-SBK & H5AN4G8NMFR-VKC  & 2666  & 4     & 8     & x8 & 1176.0 & 1382.9 & 1165.1 \\
        $M12$  & TLRD44G2666HC18F-SBK & H5AN4G8NMFR-VKC   & 2666  & 4     & 8     & x8  & 1485.0 & 1740.6 & --\\
        \stripe
        $M13$  & KSM32RD8/16HDR & H5AN4G8NAFA-UHC& 2400  & 4     & 8     & x8  & 1853.5 & 2849.6 & --  \\
        $M14$ & F4-2400C17S-8GNT & H5AN4G8NMFR-UHC & 2400  & 8     & 8     & x8   & 1369.3 & 1942.2 & --  \\
        \stripe
        $M15$  & F4-2400C17S-8GNT & H5AN4G8NMFR-UHC & 3200  & 8     & 8     & x8 & 1545.8 & 2147.2 & --  \\
        $M16$  & KSM32RD8/16HDR & H5AN8G8NDJR-XNC   & 3200  & 16     & 8     & x8 & 1634.4 & 1944.6 & -- \\
        \stripe
        $M17$ & KSM32RD8/16HDR &  H5AN8G8NDJR-XNC& 3200  &  16    & 8     & x8 & 1664.7 & 2016.6 & --  \\
        \midrule
        \multicolumn{10}{l}{$^{\dagger{}}$The maximum possible entropy in a DRAM segment is 64K (65,536) bits.}
    \end{tabular}%
    } 
    \label{table:ddr4_table}%
\end{table*}%

\end{document}